\documentclass[amsmath,amssymb,10pt,aps,notitlepage,twocolumn,floatfix]{revtex4-1} 
 \usepackage[colorlinks=true]{hyperref}
 \usepackage{lipsum}
 \usepackage{graphicx}
 \usepackage{textcomp}
\usepackage[dvipsnames]{xcolor}
\usepackage{latexsym}
\usepackage{amsmath}
\usepackage{amsthm}
\usepackage{amssymb}
\usepackage{epstopdf} 
\usepackage{enumerate}
\usepackage{setspace} %
\usepackage{dcolumn}%
\usepackage{bm}%
\usepackage{setspace} %
\usepackage{slashed}
\usepackage{color}
\usepackage{xcolor}
\usepackage{youngtab}
\usepackage{tikz}
\usepackage{tikz-3dplot}
\usepackage{braket}
\usetikzlibrary{shapes,snakes,arrows,chains,matrix,positioning,scopes,calc}
\usetikzlibrary{decorations.markings}

\usepackage[tabbotcap]{subfigure}

\allowdisplaybreaks
 
\begin{document}
\title{Realization of fractonic quantum phases in the breathing pyrochlore lattice}

\author{SangEun Han}
\thanks{These authors contributed equally to this work.}
\author{Adarsh S. Patri}
\thanks{These authors contributed equally to this work.}
\author{Yong Baek Kim}

\affiliation{Department of Physics, University of Toronto, Toronto, Ontario M5S 1A7, Canada
}

\date{\today}

\begin{abstract}
Fractonic phases of matter are novel quantum ground states supporting sub-dimensional emergent excitations with mobility restrictions. Due to a sub-extensive ground state degeneracy that is dependent on the geometry of the underlying lattice, fractonic phases are considered as models for quantum memory or quantum glass. While there exist a number of exactly solvable models with interactions between multiple particles/spins, the realization of such models in real materials is extremely challenging. In this work, we provide a realistic quantum model of quadratic spin interactions on the breathing pyrochlore lattice of existing materials. We show that the emergent ``cluster charge'' excitations arise as vacuum fluctuations residing on the boundary of membrane objects, and move in a sub-dimensional space. Using the membrane operators, we demonstrate the existence of a sub-extensive ground state degeneracy explicitly depending on the lattice geometry, which is a useful resource for novel quantum memory.
\end{abstract}

\maketitle

\section{Introduction}
From the Landau quasiparticle of Fermi liquid theory to the Majorana fermions in the Kitaev model, emergent quasiparticles provide deep insight into the nature of strongly-interacting many-body systems.
Despite the variety of settings in which they may arise, the common feature that all known quasiparticles typically share is their ability to (freely) move.
Fractonic phases of matter are a rare example that fails to conform with this conventional wisdom \cite{PhysRevLett.94.040402,doi:10.1080/14786435.2011.609152,PhysRevA.83.042330,PhysRevB.92.235136,PhysRevB.94.235157,doi:10.1146/annurev-conmatphys-031218-013604,doi:10.1142/S0217751X20300033,PhysRevResearch.2.033300,wang2020nonliquid}.
With an underlying ground state degeneracy that is sub-extensive in system size \cite{PhysRevA.83.042330,PhysRevB.92.235136,PhysRevB.95.245126,PhysRevB.94.235157,PhysRevB.95.245126,PhysRevB.96.195139,doi:10.1146/annurev-conmatphys-031218-013604,doi:10.1142/S0217751X20300033,vijay2017isotropic,10.21468/SciPostPhysCore.4.2.012,PhysRevX.8.031051}, their emergent quasiparticles come in two varieties: (i) sub-dimensional excitations such as lineons or planeons \cite{PhysRevLett.94.040402,PhysRevB.94.235157,BRAVYI2011839,PhysRevB.92.235136,10.21468/SciPostPhys.6.1.015,10.21468/SciPostPhys.6.4.043}, where the respective particles are restricted to move along particular lines or planes in a three-dimensional system, and (ii) immobile excitations, known as fractons \cite{PhysRevA.83.042330,PhysRevB.88.125122}.
The immobility of fractonic excitations is intimately linked to the notion that attempting to move them results in ``bursts'' of additional particles being created; in the absence of a constant input of energy to accommodate these additional particles, fractonic excitations are thus transfixed in space.
Though these single excitations may be forbidden to propagate seamlessly, composites formed from these elementary excitations may be free and mobile through the system.

Fracton models have recently come under intense investigation in quantum error-correcting codes, such as the X-cube model and Haah's code \cite{PhysRevA.83.042330,PhysRevB.94.235157}.
The X-cube model is composed of qubits residing on the edges of a cubic lattice \cite{PhysRevB.94.235157}, with a Hamiltonian composed of a cube operator (product of Pauli X operators residing on the twelve edges of a cube) and a cross operator (product of four Pauli Z operators in the plane touching a vertex of the cube); here the Pauli operators act on the qubit basis states.
This model supports a sub-extensive ground state degeneracy ($\log_{2}$(GSD) $\sim 2(L_{x} + L_y + L_z) - 3$, where $L_{x,y,z}$ are the system dimensions), as well as sub-dimensional and fractonic excitations \cite{PhysRevB.94.235157,PhysRevB.95.245126,vijay2017isotropic}.
Haah's code is also defined on a cubic lattice with two qubits on every vertex of the lattice \cite{PhysRevA.83.042330}; the Hamiltonian is composed of a product of a pair of Pauli operators on each vertex of the cube.
This model also possesses a sub-extensive ground state degeneracy \cite{PhysRevA.83.042330,Haah2013}, and only possesses immobile fractonic excitations. %
Importantly, the ground state degeneracy of such models is not solely dependent on the topology of the underlying manifold, but on the geometry of the lattice on which it is defined.
Fracton models have also been naturally discussed in the context of higher-rank gauge theories \cite{PhysRevB.95.115139,PhysRevB.96.035119,PhysRevB.97.235112,PhysRevLett.124.050402,10.21468/SciPostPhys.8.4.050,10.21468/SciPostPhys.10.1.003,10.21468/SciPostPhys.9.4.046,PhysRevB.97.085116,PhysRevB.96.125151,bulmash2018generalized,PhysRevResearch.2.023249,PhysRevX.9.031035,YOU2020168140,PhysRevB.100.125150}, where the conservation laws associated with the modified Gauss's law constraints leads to restricted motions of quasiparticles as well as immobile fractonic excitations. 
In particular, in symmetric vector rank-2 U(1) gauge theories, where the electric and magnetic potential are rank-2 tensorial objects with an associated vector charge $\vec{\rho}$, the conservation of ``linear momentum'' ($\vec{Q} = \int \vec{\rho}$) and ``angular momentum'' ($\vec{M} = \int \vec{\rho} \times \vec{x}$) leads to lineon excitations where particles are only permitted to move along the direction of the vector charge \cite{PhysRevB.95.115139,PhysRevB.96.035119}.

Despite the elegant nature of the corresponding low-energy descriptions and their novel properties, at the microscopic level the aforementioned lattice models \cite{PhysRevB.92.235136,PhysRevB.94.235157,YOU2020168140,
vijay2017isotropic,PhysRevB.95.155133,PhysRevX.8.031051,PhysRevB.100.125150,YOU2020168140,PhysRevB.100.235115,PhysRevB.97.165106,PhysRevB.98.035112,PhysRevResearch.2.043165} possess complicated multi-spin interactions that provide a challenging task to realize in a concrete experimental setting.
Indeed, to make further theoretical progress, it would also be beneficial to have a situation wherein fractonic and sub-dimensional excitations naturally emerge due to the geometry constraints of the system as well as the interacting nature of the microscopic objects.
Previous works have included coupled spin chain systems \cite{PhysRevLett.119.257202}, and Kitaev-type interactions on the hyper-honeycomb model \cite{PhysRevB.96.165106}. Recently it was shown that a classical spin liquid on the breathing pyrochlore lattice \cite{PhysRevLett.124.127203}, where interactions amongst classical spins residing on the vertices of two unequal corner-sharing tetrahedra, possesses a low energy manifold described by an underlying classical rank-2 vector gauge theory.
Specifically, the rank-2 electric field tensor is populated by linear combinations of spins on the four sublattices of a tetrahedron (i.e. the ``light'' normal modes of the $T_d$ point group tetrahedron, whose fluctuations are energetically inexpensive), with a corresponding Gauss's law constraint due to the suppression of energetically costly ``heavy modes''  \cite{PhysRevLett.124.127203}.
The subsequent classical ground state has been shown to exhibit fourfold pinch point singularities in certain correlation functions that may be resolved under neutron scattering experiments \cite{PhysRevB.98.165140,PhysRevLett.124.127203}. 

In this work, motivated by this classical study, we demonstrate that a quantum model on the breathing pyrochlore lattice can support a fractonic phase of matter.
In the limit of particular energy penalties associated with normal mode fluctuations, we show that the corresponding quantum theory leads to spinor charges with mobility restrictions.
We also discuss lack of local operators, except for a membrane operator where only the corners of the membrane occupied by the spinor charges are permitted to move.
We numerically discover that the ground state degeneracy is not extensive in volume, and is strongly dependent on the lattice geometry.
We further argue that a local magnetic field, which would allow the quantum system to tunnel in between the degenerate ground states, is prohibited in this geometry in the thermodynamic limit.
Indeed, magnetic field terms are only permitted under perturbations that extend to the boundary of the system, and is thus suppressed in the thermodynamic limit.
The lack of mobile excitations and a non-extensive (yet geometry dependent) ground state degeneracy strongly suggests that the breathing pyrochlore lattice supports a quantum fractonic phase of matter.

The remainder of the paper is organized as follows. 
In Sec.~\ref{sec_r2_u1_summary} we provide an overview of the important aspects of higher-rank U(1) gauge theories, and recap the classical breathing pyrochlore model within the framework of rank-2 vector gauge theory in Sec.~\ref{sec_classical_bp}.
We then present the quantum breathing pyrochlore lattice model in Sec.~\ref{sec_quantum_bp} and elucidate the 
quantum ground state degeneracy, the variety of perturbative terms, and the excitations in terms of spinor charge degrees of freedom.
We also emphasize the occurrence of a thermodynamically large membrane operator that permits these spinor excitations to be moved to the boundary of the system, and argue the prohibition of local magnetic field terms (or perturbative terms that connect the various degenerate ground states) due to the complicated geometrical configuration of our setting.
Finally, in Sec.~\ref{sec_discussions} we discuss the broad implications of our work and propose future directions of exploration.

\section{Higher-Rank U(1) Gauge Theories}\label{sec_r2_u1_summary}

The interacting classical and quantum spin models on the breathing pyrochlore lattice have an underlying higher-rank gauge theory structure that emerges in the low energy limit \cite{PhysRevLett.124.127203}.
To specify our notation and terminology, we present a succinct overview of rank-2 U(1) gauge theories in this section. 

The classical theory of electromagnetism is described in terms of a rank-1 gauge theory, wherein the electric field ($E_i$) and magnetic vector potential ($A_i$) transform as vectors under spatial rotations.
Associated with this familiar Maxwell theory is a source-free Gauss' law constraint for the electric field, $\partial_i E_i = 0$, and a U(1) gauge transformation, $A_i (x) \rightarrow A_i (x) + \partial_i \lambda(x)$, for a charge density $\rho$ and an arbitrary function $\lambda(x)$; we note that repeated indices are summed over.
This gauge transformation can be simply obtained by acting the source-free Gauss' law on a state vector/wavefunction.
At higher-energies, this source-free condition can be relaxed to $\partial_i E_i = \rho \neq 0$, which allows the creation of charges of the electric field; for a compact gauge theory (where $A_i$ is defined to modulo $2 \pi$) one necessarily admits the creation of magnetic monopoles that violate the source-free Gauss's law constraint for the magnetic field, $\partial_i B_i = 0$.
We will henceforth focus on the electric charges and refer to the electric Gauss's law constraint as merely the Gauss' law for brevity.
The Gauss' law constraint  imposes a conservation law, where $\int \rho = \int \partial_i E_i = 0$, as we integrate a total derivative over the entire volume and the fields are taken to vanish on the boundary.
Physically, this ensures that charges must be created from the vacuum so that the total charge is zero i.e. an equal number of positive and negative charges.
The classical theory can be quantized by imposing that the electric field and vector potential are canonically conjugate, $[A_i(x), E_j(y)] = i \delta_{ij} \delta(x-y)$, which leads to a low-energy description with an emergent photon of dispersion $\omega \propto k$.

A natural extension of the conventional rank-1 gauge theory, is a rank-2 theory, wherein the electric field and magnetic potential are now promoted to symmetric rank-2 tensors $E_{ij}$ and $A_{ij}$, respectively \cite{PhysRevB.95.115139,PhysRevB.96.035119,rasmussen2016stable}.
Unlike in the rank-1 theory, the electric field has the possibility of satisfying distinct source-free Gauss's law constraints: (i) $\partial_i E_{ij} = 0$ and (ii) $\partial_i \partial_j E_{ij} = 0$ where the corresponding theories are referred to as vector and scalar charge theories, respectively.
These theories may be further constrained by imposing that $E_{ij}$ is traceless.
Just as in the case of rank-1 gauge theory, these distinct Gauss' law constraints lead to distinct gauge transformations for $A_{ij}$: (i) $A_{ij} \rightarrow A_{ij} +  \partial_i \lambda_j(x) + \partial_j \lambda_i(x)$ and (ii) $A_{ij} \rightarrow A_{ij} +\partial_i \partial_j \phi(x)$ for arbitrary functions $\lambda_i(x), \phi(x)$.
The distinct Gauss's laws impose a variety of possible conservation laws.
Focussing on the vector charge theory, as will be pertinent for our work, the Gauss's law constraint can be relaxed to lead to permit the creation of vector charges, $\partial_i E_{ij} = \rho_j \neq 0$.
Associated with this are a conservation of total vector charge (``linear momentum'' ) $\int \bm{\rho} = 0$, and ``angular momentum'' $\int \bm{x} \times \bm{\rho} = 0$ \cite{PhysRevB.95.115139,PhysRevB.96.035119}.
These conservation laws place strong constraints on the number of charges that may be created from the vacuum and how they may be allowed to move.
In particular, they lead to sub-dimensional excitations, where the vector charges are restricted to move along certain lines or planes, and fractonic excitations, where particles cannot only move unless extra particles are created.
In the absence of a constant energy input to facilitate the constant creation of extra particles, these fractonic excitations thus remain immobile.
Just as the rank-1 theory, this theory can also be quantized by taking the electric and magnetic tensor potentials to be canonically conjugate, $[A_{ij}(x), E_{kl}(y)] = i \left(\delta_{ik} \delta_{jl} + \delta_{il} \delta_{jk}\right) \delta(x-y)$, which leads to a low-energy description with an emergent photon of dispersion $\omega \propto k^2$ \cite{PhysRevB.96.035119,rasmussen2016stable,PhysRevB.96.125151,PhysRevB.97.235112}.
Importantly, in the quantum theory, the electric field components commute with themselves.
We direct the reader to Ref.~\cite{PhysRevB.95.115139,PhysRevB.96.035119,rasmussen2016stable,PhysRevB.96.125151} for a comprehensive description of the other rank-2 gauge theories alluded to above.

\section{Classical Breathing Pyrochlore Lattice Model}\label{sec_classical_bp}

\begin{figure}[t]
\subfigure[]{
\includegraphics[width=0.48\linewidth]{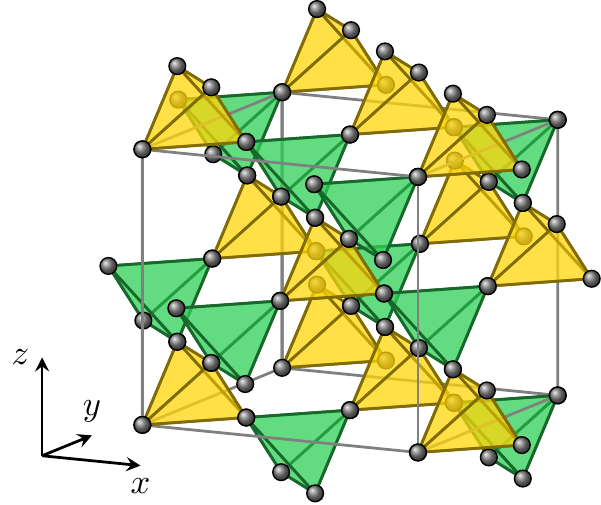}}
\subfigure[]{
\includegraphics[width=0.48\linewidth]{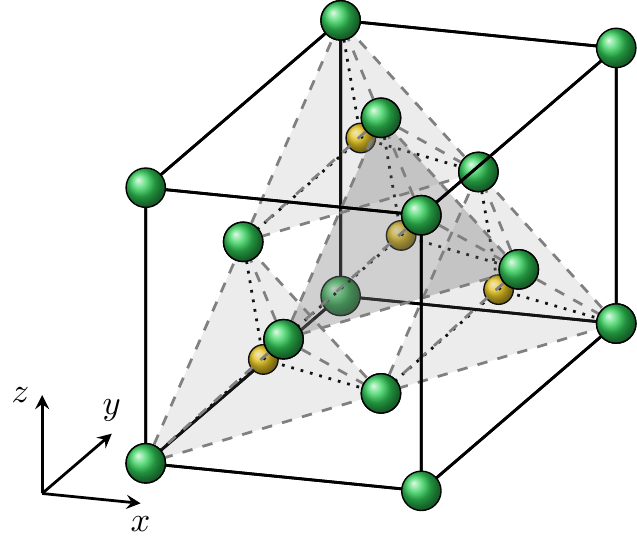}\label{fig:FCCunit}}
\caption{Breathing pyrochlore lattice. (a) The green and yellow tetrahedra stand for the A and B-tetrahedra, respectively. (b) Unit cell of the face-centered cubic lattice of A-tetrahedron ($L_{x}=L_{y}=L_{z}=1$). Each A and B-tetrahedra form the face-centered cubic lattices. The green and yellow circles stand for the center of the A and B-tetrahedra, respectively. Note that we have two planes (that contain A sites) in each direction per one unit cell.}
\label{fig_breathing_pyrochlore}
\end{figure}

The breathing pyrochlore lattice is composed of corner sharing tetrahedra of two different sizes A/B, with interactions between neighbouring spins residing on the vertices of the tetrahedra as seen in Fig.~\ref{fig_breathing_pyrochlore}.
The microscopic interactions between the spins may involve antiferromagnetic Heisenberg ($J_{\text{A}/\text{B}}$), bond-dependent Dzyaloshinskii-Moriya (DM, $D_{\text{A}/\text{B}}$) interactions \cite{PhysRevLett.124.127203}, as well as Kitaev and Gamma interactions (see Appendix~\ref{app:normalcoeff}).
The classical Hamiltonian describing the interactions between the neighbouring spins can be captured in terms of the irreducible representation formed by the spins belonging to each of the tetrahedra \cite{McClarty_2009,Benton2016,PhysRevB.95.094422,PhysRevLett.124.127203},
\begin{align}
H =  \frac{1}{2} \sum_{\text{A}, \Gamma} a_{\text{A}, \Gamma} m_{\text{A},\Gamma}^2 +  \frac{1}{2} \sum_{\text{B}, \Gamma} a_{\text{B}, \Gamma} m_{\text{B},\Gamma}^2,
\label{eq_bp_start}
\end{align}
where $\Gamma = \{\textsf{A}_{2}, \textsf{E}, \textsf{T}_2, \textsf{T}_{1+}, \textsf{T}_{1-} \}$ is over the $T_d$ irreps for a given tetrahedron (A or B), $a_{\text{A}/\text{B}, \Gamma}$ are the interaction coefficients, and $m_{\text{A}/\text{B},\Gamma}$ denotes the pseudospin corresponding to different irrep on the A/B tetrahedron.
Microscopically (as presented in Appendix~\ref{app:normalcoeff}), one can minimally take antiferromagnetic $J_\text{A}, J_\text{B} >0$, while taking $D_\text{A} < 0$, and $D_\text{B} = 0$. 
With this choice, on the B-tetrahedron, $a_{\text{B}, \textsf{T}_{1+}} > 0$, while the remaining modes are negative \cite{PhysRevLett.124.127203}. %
As a consequence, at low energies, the fluctuations of the $\textsf{T}_{\text{B},1+}$ mode are energetically costly leading to $\mathbf{m}_{\text{B}, \textsf{T}_{1+}} = 0$.
Analogously, for the A-tetrahedron, the DM interaction leads to having small (and negative) interaction coefficients for a number of interaction coefficients. However, since still $a_{\text{A},\text{T}_{1+}}>0$, it energetically leads to $\textbf{m}_{\text{A},\text{T}_{1+}}=0$, as shown in Appendix~\ref{app:normalcoeff}. 

\begin{figure}
\includegraphics{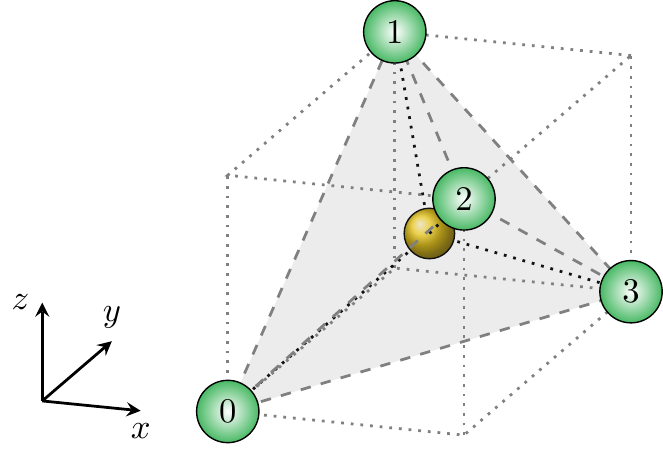}
\caption{The location of the A-tetrahedra surrounding the B-tetrahedron. Here, the yellow circle stands for the B-tetrahedron and the green circles stand for the A-tetrahedra surrounding B-tetrahedron.}\label{fig:convention}
\end{figure}

This constraint on $\mathbf{m}_{(\text{A},\text{B}), \textsf{T}_{1+}}$ can be rewritten in terms of the normal modes of the surrounding four A tetrahedron, as seen in Fig.~\ref{fig:convention}.
Performing a gradient expansion of the A normal modes about the central B site location (as described in Appendix \ref{app:gausscontinu}), we arrive at the continuity equation,
\begin{align}
& \frac{2}{\sqrt{3}} \begin{pmatrix}
\partial_x m_{\text{A}, \textsf{E}}^1 \\
-\frac{1}{2}\partial_y m_{\text{A}, \textsf{E}}^1 + \frac{\sqrt{3}}{2} \partial_y m_{\text{A}, \textsf{E}}^2 \\
-\frac{1}{2}\partial_z m_{\text{A}, \textsf{E}}^1 - \frac{\sqrt{3}}{2} \partial_z m_{\text{A}, \textsf{E}}^2
\end{pmatrix} 
 + \begin{pmatrix}
\partial_y m_{\text{A}, \textsf{T}_{1-}}^z + \partial_z m_{\text{A}, \textsf{T}_{1-}}^y \\
\partial_x m_{\text{A}, \textsf{T}_{1-}}^z + \partial_z m_{\text{A}, \textsf{T}_{1-}}^x \\
\partial_x m_{\text{A}, \textsf{T}_{1-}}^y + \partial_y m_{\text{A}, \textsf{T}_{1-}}^x
\end{pmatrix} \nonumber \\
& - \sqrt{\frac{2}{3}} \bm{\nabla} {m}_{\text{A}, \textsf{A}_2}  - \mathbf{\nabla} \times \mathbf{m}_{\text{A}, \textsf{T}_2}%
= 0 .
\label{eq_continuity}
\end{align}

The combination of the normal modes of a single $A$ tetrahedron can be expressed in terms of a rank-2 tensor, $\mathbf{E}_{\text{A}} = \mathbf{E}_{\text{A}} ^{\text{sym}} +  \mathbf{E}_{\text{A}} ^{\text{antisym}} +  \mathbf{E}_{\text{A}} ^{\text{trace}}$, where we have suggestively decomposed into a symmetric tensor,

\begin{align}
&\mathbf{E}_{\text{A}} ^{\text{sym}} \notag\\
&=\left(\begin{matrix}
\tfrac{2}{\sqrt{3}}m_{\text{A},\textsf{E}}^{1} & m_{\text{A},\textsf{T}_{1-}}^{z} & m_{\text{A},\textsf{T}_{1-}}^{y}\\
m_{\text{A},\textsf{T}_{1-}}^{z} & -\tfrac{1}{\sqrt{3}}m_{\text{A},\textsf{E}}^{1}+m_{\text{A},\textsf{E}}^{2} & m_{\text{A},\textsf{T}_{1-}}^{x}\\
m_{\text{A},\textsf{T}_{1-}}^{y} & m_{\text{A},\textsf{T}_{1-}}^{x} & -\tfrac{1}{\sqrt{3}}m_{\text{A},\textsf{E}}^{1}-m_{\text{A},\textsf{E}}^{2}\\
\end{matrix}\right),
\end{align}
anti-symmetric tensor, $(\mathbf{E}_{\text{A}} ^{\text{antisym}})_{ij} = - \epsilon_{ijk} m_{\text{A}, \textsf{T}_2}^k$, and a traceful tensor $(\mathbf{E}_{\text{A}} ^{\text{trace}})_{ij} = -\delta_{ij} \sqrt{\frac{2}{3}} m_{\text{A},A_2}$. 
The formulation of the normal modes in terms of the rank-2 electric field tensor allows one to notice that the electric field tensor satisfies the Gauss' law for rank-2 vector gauge theory, $\nabla\cdot\mathbf{E}_{\text{A}}=0$ \cite{PhysRevB.96.165106}.

Evidently, the continuity equation involves a number of A-tetrahedron irreps in non-trivial combinations.
To shine light on the underlying structure, we recall that though the interaction coefficients may involve the aforementioned microscopic coupling parameters, from group-theoretic methods the interaction coefficients are merely known to be (in general) distinct from each other \textit{a priori}.
Indeed, one would naturally expect that including further neighbour-interactions, for instance, may renormalize these interaction coefficients.
To that end, we consider the case where $a_{\textsf{A}_{2}} = a_{\textsf{E}} < 0$ on the A-tetrahedron, and take the remaining A-modes to be energetically positive and costly (we explicitly show such a microscopic construction in Appendix~\ref{app:normalcoeff}).
In a similar fashion to the B-tetrahedron normal modes, this leads to only $m_{\text{A},\textsf{A}_{2}}, \mathbf{m}_{\text{A},\textsf{E}} \neq 0$ on the A-tetrahedron.
As will be seen in the next section, the simple choice of the interaction coefficients is necessary to ensure a closed algebra for the normal modes in the quantum breathing pyrochlore model.
Using the defined electric field tensors, the equation $\nabla\cdot(\mathbf{E}_{\text{A}} ^{\text{sym}} +  \mathbf{E}_{\text{A}} ^{\text{antisym}} 
+ \mathbf{E}_{\text{A}} ^{\text{trace}})=0$ then takes an elegant form,
\begin{align}
\partial_i \Big[ \mathbf{E}_{\text{A}} ^{\text{sym}}+\mathbf{E}_{\text{A}} ^{\text{trace}} \Big]_{ii} = \mathbf{0}, \ \forall i \in \{x,y,z\},
\label{eq_gauss_law}
\end{align}
which is identical to the Gauss's law constraint for a rank-2 gauge theory for a vector charge density $\bm{\rho} =0$.
Evidently, the classical (microscopic) breathing pyrochlore lattice model has an emergent classical rank-2 vector gauge theory description.
We note that Eq.~\ref{eq_gauss_law} holds even with non-vanishing $\textbf{m}_{\text{A},\textsf{T}_{1-}}$ normal mode.
We henceforth define $(\mathbb{E}_{\text{A}})_{ij} = -\sqrt{2}(\mathbf{E}_{\text{A}} ^{\text{sym}}+\mathbf{E}_{\text{A}} ^{\text{trace}})_{ij} $ (where we mutiply it by $-\sqrt{2}$ for later convenience), and the source-free Gauss's law constraint becomes $\partial_i (\mathbb{E}_{\text{A}})_{ii} = \bm{0} \  \forall i \in \{x,y,z\}$.

\section{Quantum Breathing Pyrochlore Lattice Model} \label{sec_quantum_bp}

The quantum breathing pyrochlore model involving the $\textsf{A}_2$ and $\textsf{E}$ irrep normal modes can be written as $H = H_0 + H'$, where
\begin{align}
H_0 = - 4| a_\text{A} | \sum_{\text{A}} \left( \mathbf{m}_{\text{A}, \textsf{E}}^2 + m_{\text{A},\textsf{A}_2}^2 \right ),
\end{align}
with $a_{\text{A},\textsf{A}_2} = a_{\text{A},\textsf{E}} = -  8|a_\text{A}|$, and 
\begin{align}
H' = & \frac{1}{2} \sum_{\text{B}, \Gamma} a_{\text{B}, \Gamma} m_{\text{B},\Gamma}^2.
\label{eq_h_prime_pert}
\end{align}
In the quantum model, the electric field components satisfy a canonically normalized SU(2) Lie algebra, $[ \mathbb{E}_{\text{A}, i}, \mathbb{E}_{\text{A}', j} ] = i \delta_{\text{A},\text{A}'} \epsilon_{i j k}  \mathbb{E}_{\text{A}, k}$, where $\{i,j,k \} \in \{xx, yy, zz \}$ (see Appendix~\ref{app:normalcoeff}).
Note that the electric field variables do not commute. %
With these  electric field variables, $H_0$ takes the simple form, 
\begin{align}
H_0 & = - | a_\text{A} | \sum_\text{A} \left( \mathbb{E}_{\text{A},xx}^2 + \mathbb{E}_{\text{A},yy}^2 + \mathbb{E}_{\text{A},zz} ^2 \right) \nonumber \\
& = - | a_\text{A} | \sum_\text{A} \vec{\mathbb{E}}_{\text{A}} ^2 \label{eq_h_0_quantum},
\end{align}
as $m_{\text{A},\textsf{A}_{2}}$ and $\mathbf{m}_{\text{A},\textsf{E}}$ can be written in terms of the diagonal components of the electric field tensor as mentioned in the previous section (also see Appendix~\ref{app:normalcoeff}).
To make progress, we make a choice for the remaining B tetrahedron normal modes interaction coefficients.
In particular, we take the $a_{\text{B}, \Gamma}$ coefficients to be perturbatively small as compared to the A tetrahedron coefficients.
Such a choice is certainly permitted as the A and B tetrahedron possess their own microscopic interactions, and permits a controlled study of the low-energy description of the quantum model. Indeed, in the quantum spin ice model on the pyrochlore lattice \cite{PhysRevB.69.064404,PhysRevB.94.075146}, the quantum flip terms $J_{\pm} S^{+} S^{-}$ were taken to be perturbatively weaker than the Ising interaction $J_{zz}$ between neighbouring spins. Such a choice enabled a controlled emergence of the underlying U(1) gauge structure of the model.

Diagonalizing this Hamiltonian over each (decoupled) A tetrahedron, results in an eigen-spectrum of $-6$ with five-fold degeneracy, $-2$ with nine-fold degeneracy, and $0$ with two-fold degeneracy.
Drawing inspiration from the fact that Eq.~\ref{eq_h_0_quantum} is the form of a spin-Hamiltonian $\sim \hat{S}^2$ which has a spectrum of $S(S+1)$ and corresponding degeneracy of $2S+1$, we are able to identify $-6$ eigenvalue state as corresponding to a pseudospin $S=2$ manifold of states, $-2$ eigenvalue state as corresponding to three sets of pseudospin $S=1$ manifold of states, and $0$ eigenvalue state as corresponding to two sets of pseudospin $S=0$ manifold of states.
Importantly, due to $[ {\mathbb{E}}_{\text{A}}^2, \mathbb{E}_{\text{A},zz} ] =0$, just as in typical spin algebra, we can label the states on each A-tetrahedron as $\ket{S,S^z}$, where $S (S + 1 ) \equiv {\mathbb{E}}_{\text{A}}^2$. Hence, the ground state manifold of the A-tetrahedron network can be described by $S=2$ multiplet in the low energy limit.

Relaxing the Gauss's law constraint in Eq.~\ref{eq_gauss_law} to permit the existence of charges, allows the electric charge density about a B-tetrahedron centre to be similarly defined as,
\begin{align}
{\rho}_{\text{B}} ^k = \sum_{A=0}^{3} c_{A}^{k} \mathbb{E}_{A,kk},\label{eq:rhoB}
\end{align}
where $c_A ^k$ is a site-dependent phase factor vector: $c_A ^x = (-1,-1,1,1)$, $c_A ^y= (-1,1,-1,1)$, and $c_A^z = (-1,1,1,-1)$.
The components of the vector charge density also satisfy a canonically normalized SU(2) Lie algebra, $[{\rho}_\text{B} ^i, {\rho}_\text{B} ^j] = i \epsilon_{ij}^k {\rho}_\text{B} ^k$, and as such a given state can at most be associated as the eigenstate of one of the components; we take the ${\rho}_\text{B} ^z$ eigenvalue as the label.
We note that there is not an inherently special reason for choosing the $z$-component to label the states; one can alternatively choose the $x$- or $y$- components, just as one may do so when labelling spin states in typical spin-1/2 problems.
In that sense, a given charge configuration is not represented by the value of all of it $\rho^{x,y,z}$ components, which is unlike the classical rank-2 U(1) gauge theory described in Sec. \ref{sec_r2_u1_summary}.
Thus, these charges should be considered as spinor charges.
We emphasize that the electric field variables exist on the centre of the A-tetrahedra, while the electric charges reside on the centre of the B-tetrahedra. Note that  $\rho_{\text{B}}^{z}$ have integer eigenvalues from $-8$ to $+8$ because the allowed eigenvalue of $\mathbb{E}_{zz}$ is from $-2$ to $+2$ and $\rho_{\text{B}}^{z}$ is the linear combination of $\mathbb{E}_{zz}$ on the surrounding $A$-tetrahedra.

The creation of a $\rho_\text{B} ^z$ charge from the vacuum is energetically costly.
This penalty is accounted for by the $m_{\textsf{T}_{1+}}^z$ term in Eq.~\ref{eq_h_prime_pert},
\begin{align}
\frac{1}{2} \sum_{\text{B}} a_{\text{B}, \textsf{T}_{1+}} (m_{\text{B},\textsf{T}_{1+}}^z)^2 = \frac{1}{128} \sum_{\text{B}} a_{\text{B}, \textsf{T}_{1+}} (\rho_{\text{B}} ^z)^2,
\end{align} 
where $a_{\text{B}, T_{1+}}  > 0$.
This penalty cost lifts the degeneracy of states formed by taking combinations of the pseudospin $S=2$ states on all the A-tetrahedra, and permits the ground state to be categorized with $\rho_\text{B} ^z = 0$ on every B-tetrahedron.

\subsection{Ground states: degenerate manifold}
The charge-neutral configuration is a description of the ground states.
As an illustrative example, for a given charge-neutral B tetrahedron, the states of the surrounding four A tetrahedra must conspire in a manner that satisfies the Gauss's law constraint; this can be considered as a single Gauss's law unit.
Considering the state ($\psi$) of the surrounding A tetrahedron, and imposing the charge-neutral configuration requires,
\begin{align}
&\ket{\psi} = \sum_{a,b,c,d} \mathcal{F}_{a,b,c,d} \ket{2,a}_0 \ket{2,b}_1 \ket{2,c}_2 \ket{2,d}_3 \\ 
&\implies \bra{\psi ' } (\rho_\text{B} ^z)^2 \ket{\psi} \propto \delta_{\psi, \psi'} | \mathcal{F}_{a,b,c,d} | ^2 (a-b-c+d)^2 = 0.\label{eq:gausscon}
\end{align}
where $\mathcal{F}_{a,b,c,d}$ is a complex coefficient, the subscript on the ket labels the A tetrahedron, and $\{a,b,c,d\} \in \{\pm 2,\pm 1,0\}$. The sign structure of Eq.~\ref{eq:gausscon} comes from $c_{A}^{z}=(-1,1,1,-1)$ in Eq.~\ref{eq:rhoB}.
The charge-neutral configuration thus corresponds to 85 possible states (listed in Appendix~\ref{app_charge_less_states}) for a single Gauss's law unit, thus providing a manifold of ground states.

\subsection{Variety of perturbation terms acting on degenerate ground state manifold}

\begin{figure}[t]
\subfigure[]{
\includegraphics[width=0.48\linewidth]{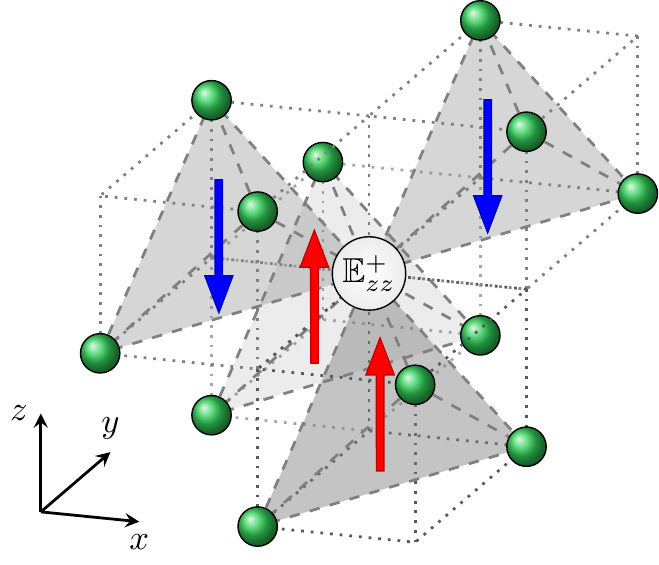}}
\subfigure[]{
\includegraphics[width=0.48\linewidth]{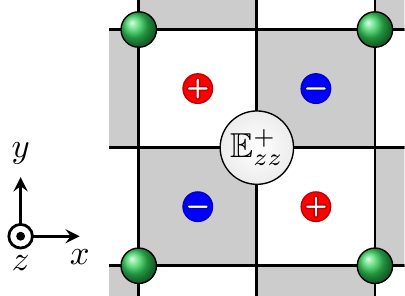}}
\caption{(a) The charge configuration when we increase $\mathbb{E}_{zz}$. The red upward and blue downward arrows stand for  $\pm1$ $z$-charges, respectively. The green circles stand the A-tetrahedra. (b) The top-down view of the charge configuration when we increase $\mathbb{E}_{zz}$ on the A-tetrahedron. The red and blue circles stand for $\pm1$ $z$-charges, respectively.}\label{fig:charge_config}
\label{fig_e_p_perturb}
\end{figure}

The charge-neutral configuration corresponds to a large  degenerate manifold of ground states.
The collection of normal modes in Eq.~\ref{eq_h_prime_pert} can act as a series of raising and lowering operators of $\mathbb{E}_{A,zz} ^{\pm} = (\mathbb{E}_{A,xx} \pm i \mathbb{E}_{A,yy})/2$, that satisfy the algebra $[ \mathbb{E}_{A,zz}, \mathbb{E}_{A,zz} ^{\pm}    ] = \pm \mathbb{E}_{A,zz} ^{\pm}  $, and importantly $[ \rho_{\text{B}}^z,  \mathbb{E}_{A,zz} ^{\pm}    ] = \pm c_{A}^z \mathbb{E}_{A,zz} ^{\pm}  $.
The second commutator results in raising and lowering the $\rho^z$ eigenvalue of the state of the system.
Figure \ref{fig_e_p_perturb} depicts the operation of the raising operator on the vacuum.
Since we take the form of the interaction coefficients in Eq.~\ref{eq_h_prime_pert} to be \textit{a priori} independent, we focus on the types of terms that may occur: (i) $ \mathbb{E}_{A,zz} \mathbb{E}_{A',zz} ^{\pm}$, (ii) $ \mathbb{E}_{A,zz}^\pm \mathbb{E}_{A',zz} ^{\pm} $, and (iii) $ \mathbb{E}_{A,zz}^\pm \mathbb{E}_{A',zz} ^{\mp} $.
Here, $A$ and $A'$ may be on the same or different A-tetrahedron location.
The perturbative terms can thus be rewritten in terms of these raising/lowering operators,
\begin{align}
H'=&\sum_{A,A'}a_{AA'}\mathbb{E}_{A,zz}\mathbb{E}_{A',zz}+\sum_{A,A'}(b_{AA'}\mathbb{E}_{A,zz}^{+}\mathbb{E}_{A',zz}^{-}+\text{h.c.})\notag\\
&+\sum_{A,A'}(c_{AA'}\mathbb{E}_{A,zz}\mathbb{E}_{A',zz}^{+}+\text{h.c.})\notag\\
&+\sum_{A,A'}(d_{AA'}\mathbb{E}_{A,zz}^{+}\mathbb{E}_{A',zz}^{+}+\text{h.c.})\label{Hb_pert}
\end{align}
where $A,A'=0,1,2,3$ represents the location of A-tetrahedron relative to B-tetrahedron on which the operator acts (Fig.~\ref{fig:convention}), and we use the generalized variables $a_{AA'},b_{AA'},c_{AA'}$ and $d_{AA'}$ that are functions of microscopic variables.
As an example, consider the $m_{\textsf{A}_{2},\text{B}}$ that can be expressed as 
$m_{\textsf{A}_{2},\text{B}}=\frac{1}{4}\sum_{A=0}^{3}m_{A,\textsf{A}_{2}}=\frac{1}{8\sqrt{3}}\sum_{A}(\mathbb{E}_{A,xx}+\mathbb{E}_{A,yy}+\mathbb{E}_{A,zz})
=\frac{1}{8\sqrt{3}}\sum_{A}(\sqrt{2}(p^{+}\mathbb{E}_{A,zz}^{-}+p^{-}\mathbb{E}_{A,zz}^{+})+\mathbb{E}_{A,zz})$  with $p^{\pm}=e^{\pm\frac{i\pi}{4}}$. The subsequent square of the aforementioned normal mode is,
\begin{align}
\notag m&_{\textsf{A}_{2},\text{B}}^{2}\\
\notag=&\frac{1}{192}\sum_{A,A'}(\sqrt{2}(p^{+}\mathbb{E}_{A,zz}^{-}+p^{-}\mathbb{E}_{A,zz}^{+})+\mathbb{E}_{A,zz})\\
\notag&\quad\quad\quad\quad\times(\sqrt{2}(p^{+}\mathbb{E}_{A',zz}^{-}+p^{-}\mathbb{E}_{A',zz}^{+})+\mathbb{E}_{A',zz})\\
\notag=&\frac{1}{96}\sum_{A,A'}(\mathbb{E}_{A,zz}^{+}\mathbb{E}_{A',zz}^{-}+\text{h.c.})+\frac{1}{96}\sum_{A,A'}(i\mathbb{E}_{A,zz}^{-}\mathbb{E}_{A',zz}^{-}+\text{h.c.})\\
\notag+&\frac{1}{192}\sum_{A,A'}\mathbb{E}_{A,zz}\mathbb{E}_{A',zz}+\frac{1}{96\sqrt{2}}\sum_{A,A'}(p^{-}\mathbb{E}_{A,zz}\mathbb{E}_{A',zz}^{+}+\text{h.c.})\\
+&\frac{1}{96\sqrt{2}}\sum_{A,A'}(p^{-}\mathbb{E}_{A,zz}^{+}\mathbb{E}_{A',zz}+\text{h.c.}).
\end{align}
The details about representing the normal modes on B-tetrahedron in terms of the normal modes on the surrounding A-tetrahedra can be found in Appendix~\ref{app:BtoA}.
Note that $\mathbb{E}_{A,zz}^{a}\mathbb{E}_{A',zz}^{b}$ ($a,b=\pm,1$) acts on different planes, depending on the locations of $A,A'$, as shown in Table~\ref{tab:int_plane}.

\subsection{Membrane operators from perturbation}

\begin{table}
\begin{tabular}{|>{$}c<{$}|>{$}c<{$}|}
\hline
\hline
(A,A')&\text{plane}\\
\hline
(0,3),(1,2)&xy\\
(0,2),(1,3)&xz\\
(0,1),(2,3)&yz\\
\hline
\hline
\end{tabular}\caption{The planes in which $\mathbb{E}_{A,zz}^{a}\mathbb{E}_{A',zz}^{b}$ ($a,b=\pm,1$) acts depending on $A,A'$. For example, for $(A,A')=(1,2)$, $\mathbb{E}_{1,zz}^{a}\mathbb{E}_{2,zz}^{b}$ acts on $xy$-plane.}\label{tab:int_plane}
\end{table}

\begin{figure}
\centering
\includegraphics{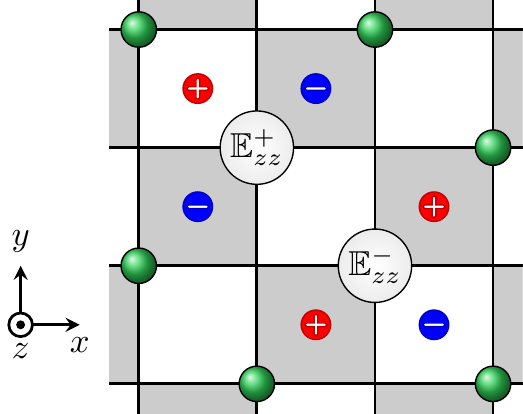}
\caption{The top-down view of charge configuration for $\mathbb{E}_{1,zz}^{+}\mathbb{E}_{2,zz}^{-}$. Here, the gray and white squares represent the positions of the charges located on $z=1/4$ and $z=-1/4$, respectively. The green circles indicate the location of the A-tetrahedra on the same $xy$-plane, $z=0$.}
\label{fig:perturb_unit}
\end{figure}

\begin{figure*}
\subfigure[]{
\includegraphics{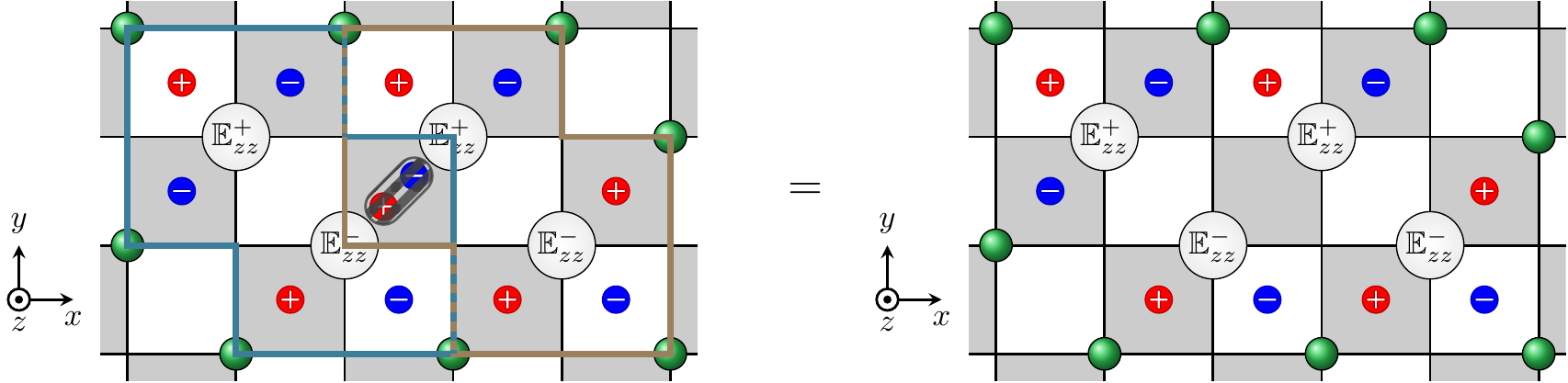}\label{fig:perturb_2}}
\subfigure[]{
\includegraphics{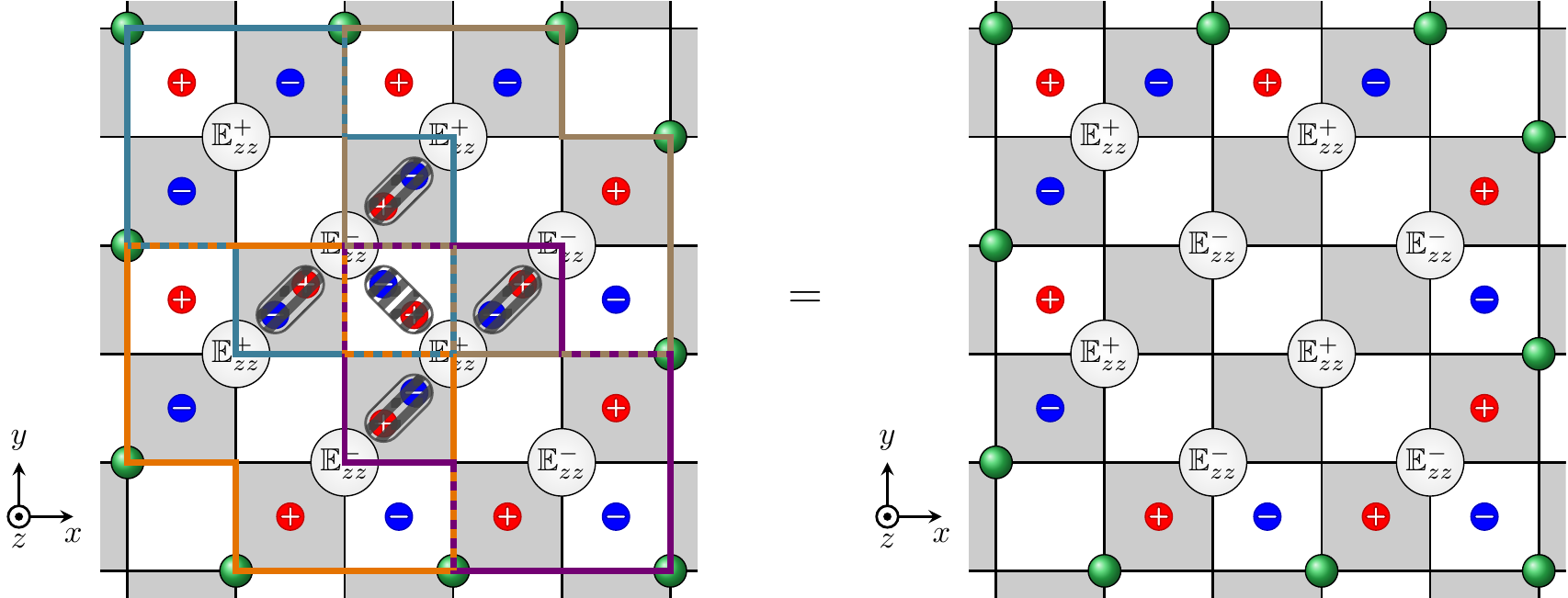}\label{fig:perturb_4}}
\caption{Depiction of the perturbation by $\mathbb{E}_{1,zz}^{+}\mathbb{E}_{2,zz}^{-}$. (a) is for the second order perturbation, and (b) is for the fourth order perturbation. The charges in the overlapped region are cancelled out, but there are still remaining charges on the edge.}
\end{figure*}

\begin{figure}
\includegraphics[width=0.8\linewidth]{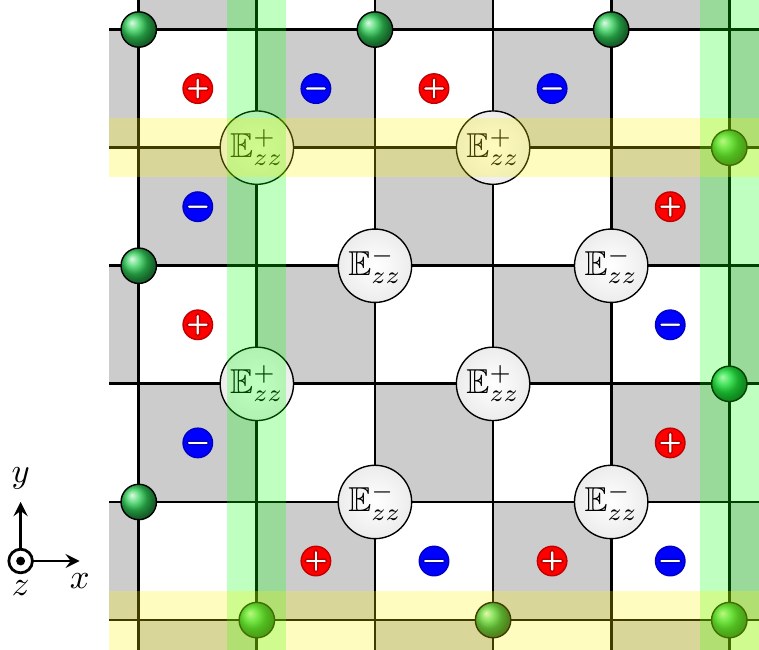}
\caption{The periodic boundary condition to cancel out the charges on the edge of the membrane operators. By identifying green lines, the charges on the left side of the green lines are cancelled out. Similarly, by identifying yellow lines, the charges above the yellow lines are cancelled out. As a result, we can get another charge-neutral state. }\label{fig:identify}
\end{figure}

The perturbative terms permit the construction of a \textit{membrane} operator that allows charges created from the vacuum to be brought to the boundaries of the system. 
As an illustrative example, we focus on $\mathbb{E}_{1,zz}^{+}\mathbb{E}_{2,zz}^{-}$, and perform degenerate perturbation theory on the ground states on the same $xy$ plane. 
At first-order in perturbation, this results in a state that has the charge configuration presented in Fig.~\ref{fig:perturb_unit}. 
This is in fact an excited state, due to the presence of non-trivial charges, and as such there is no overlap with the underlying (charge-neutral) ground state manifold. %
At second-order in perturbation on the same $xy$-plane, the leading contribution arises from the charge combinations presented in Fig.~\ref{fig:perturb_2}. 
In this case, the charges residing on the overlapping regions between the successive $\mathbb{E}_{1,zz}^{+}\mathbb{E}_{2,zz}^{-}$ operations are cancelled out, leaving behind a charge on the `edge'. 
Repeating the application of the perturbative $\mathbb{E}_{1,zz}^{+}\mathbb{E}_{2,zz}^{-}$ term on the same $xy$-plane, one obtains (at higher-orders in perturbation) a leading order contribution of charges that resemble a \emph{membrane}. 
Importantly, there is no charge inside the membrane, due to the aforementioned cancellation, and the remaining charges reside on the edge (Fig.~\ref{fig:perturb_4}); as such this is still an excited state. 
However, by imposing the appropriate periodic boundary conditions, these edge charges may be cancelled out. %
For example, in Fig.~\ref{fig:identify}, by identifying green and yellow lines as the adjoining boundaries, the positive and negative edge charges are promptly cancelled out.  %
We have thus returned back to the charge-neutral vacuum, which is distinct from the original charge-neutral ground state due to the application of the raising/lowering operators that have given different electric field quantum numbers on A-tetrahedron sites.
In this sense, the application of the membrane operator, in conjunction with the appropriate boundary condition, makes the quantum system to be able to tunnel between its manifold of ground states.
 
\begin{figure*}
\includegraphics{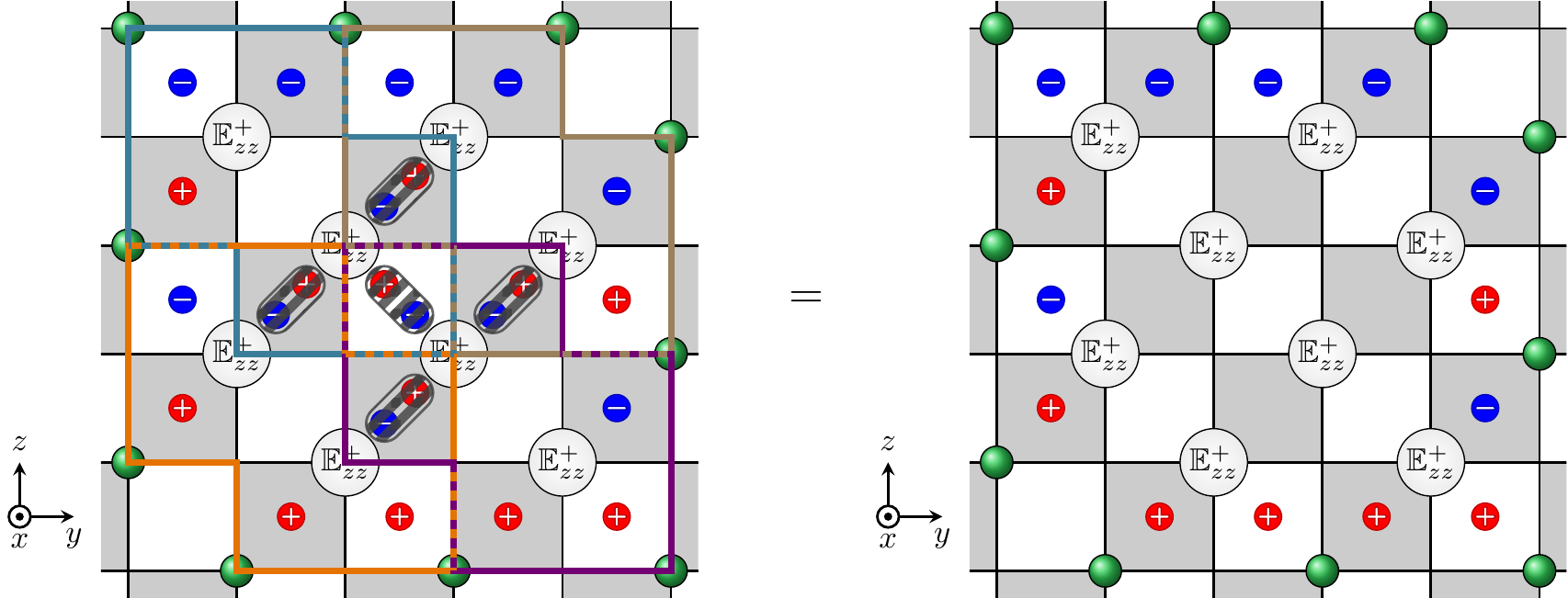}
\caption{Depiction of the fourth-order perturbation by $\mathbb{E}_{2,zz}^{+}\mathbb{E}_{3,zz}^{+}$ on $yz$ plane. When we assume that $\mathbb{E}_{2,zz}^{+}\mathbb{E}_{3,zz}^{+}$ acts on $yz$ plane at $x=0$, the gray and white squares  represent the positions of the charges located on $x=1/4$ and $x=-1/4$. 
The charges in the overlapped region are cancelled out, but there are still remaining charges on the edge.}\label{fig:yz_pert4}
\end{figure*}

\begin{figure}
\includegraphics{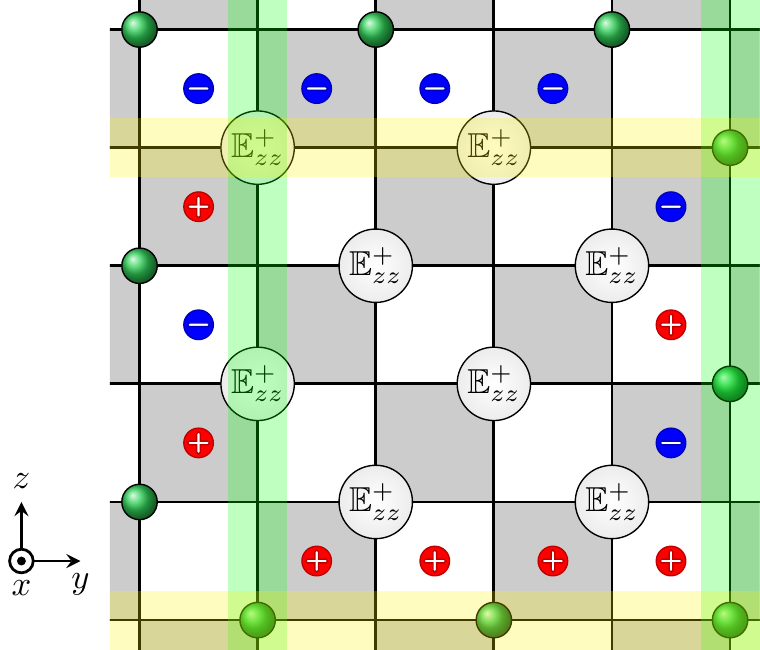}
\caption{The periodic boundary condition to cancel out the charges on the edge of the membrane operators consisting of $\mathbb{E}_{2,zz}^{+}\mathbb{E}_{3,zz}^{+}$ on $yz$ plane. By identifying green lines, the charges on the left side of the green lines are cancelled out. Similarly, by identifying yellow lines, the charges above the yellow lines are cancelled out. As a result, we can get another charge-neutral state. }\label{fig:yz_pbc}
\end{figure}

Similarly, we may be able to construct membrane operators on each of the cubic planes of the system.
The membrane operators on $xy$-plane are generated by $\mathbb{E}_{1,zz}^{\pm}\mathbb{E}_{2,zz}^{\mp}$ and $\mathbb{E}_{0,zz}^{\pm}\mathbb{E}_{3,zz}^{\mp}$, the membrane operators $xz$-plane are generated by $\mathbb{E}_{0,zz}^{\pm}\mathbb{E}_{2,zz}^{\pm}$ and $\mathbb{E}_{1,zz}^{\pm}\mathbb{E}_{3,zz}^{\pm}$, and the membrane operators $yz$-plane are generated by $\mathbb{E}_{0,zz}^{\pm}\mathbb{E}_{1,zz}^{\pm}$ and $\mathbb{E}_{2,zz}^{\pm}\mathbb{E}_{3,zz}^{\pm}$, respectively. 
For example, by using $\mathbb{E}_{2,zz}^{+}\mathbb{E}_{3,zz}^{+}$ we can make perturbation on $yz$ plane shown in Fig.~\ref{fig:yz_pert4}, and taking periodic boundary conditions in the $y$ and $z$ directions, we can return back to the charge-neutral vacuum (Fig.~\ref{fig:yz_pbc}).

We note that if we were to use a different set of operators such as $\mathbb{E}_{1,zz}^{\pm}\mathbb{E}_{2,zz}^{\pm}$ instead of $\mathbb{E}_{1,zz}^{\pm}\mathbb{E}_{2,zz}^{\mp}$, 
we would need to apply it on all the A-tetrahedron sites in the entire lattice, not just on the plane, in order to return back to the charge-less vacuum (having imposed the appropriate periodic boundary conditions) as discussed in Appendix~\ref{app:bulk_op}. 
However, the action of these stacked operators can be replicated by taking combinations of the membrane operators consisting of $\mathbb{E}_{0,zz}^{\pm}\mathbb{E}_{2,zz}^{\pm}\text{ and }\mathbb{E}_{1,zz}^{\pm}\mathbb{E}_{3,zz}^{\pm}$ on $xz$ planes, or $\mathbb{E}_{0,zz}^{\pm}\mathbb{E}_{1,zz}^{\pm}\text{ and }\mathbb{E}_{2,zz}^{\pm}\mathbb{E}_{3,zz}^{\pm}$ on $yz$ planes; this is discussed in more depth in Appendix~\ref{app:bulk_op}.
Thus, the aforementioned membrane operators are the fundamental operators on the breathing pyrochlore lattice. 

Here, we have some remarks. Firstly, we discussed the membrane operators from $\mathbb{E}_{i,zz}^{\pm}\mathbb{E}_{j,zz}^{\pm}$ and $\mathbb{E}_{i,zz}^{\pm}\mathbb{E}_{j,zz}^{\mp}$ above. 
However, from Eq.~\ref{Hb_pert}, we recall that there exist other terms in the perturbative Hamiltonian, namely $\mathbb{E}_{i,zz}\mathbb{E}_{j,zz}$ and $\mathbb{E}_{i,zz}\mathbb{E}_{j,zz}^{\pm}$. 
Importantly, these other operators does not yield any nontrivial results. 
For instance, when we act $\mathbb{E}_{i,zz}\mathbb{E}_{j,zz}$, we are in the same ground state because it does not change given state,
while $\mathbb{E}_{i,zz}\mathbb{E}_{j,zz}^{\pm}$ does yield charge-ful excited states. 
If we allow the operators to act on the same sites twice, we can make the membrane operators by using $\mathbb{E}_{i,zz}\mathbb{E}_{j,zz}^{\pm}$.
This however requires a higher order of perturbation than the membrane operators discussed above. 
Secondly, when we constructed the membrane operators, we apply the same perturbative operators on the site on the plane. 
If we used a mixture of the several types of perturbative operators in Eq.~\ref{Hb_pert}, 
we may get the charge-ful excited states, and even if we apply periodic boundary conditions, so we are unable to return back to the charge-neutral ground states.
We provide an illustrative example and discussion in Appendix~\ref{app:mix_op}.

\subsubsection{Subsystem symmetry of membrane operators}

The application of the membrane operator results in the creation of charges that obey certain conservation laws on the planes on which the operator is acting on.
Indeed, the charges are independently conserved on each of the planes. 
This arises due to the charge configuration created when we increase or decrease the electric fields on the A-tetrahedron (Fig.~\ref{fig:charge_config}).
Thus the creation of the charges on the various planes is a consequence of not only conserving the total charge, but also the charge within a plane as well. 
This emergent subsystem symmetry can be interpreted as the conservation ``first moment'' of the charge in the plane $\sum_i y_i \rho_z = 0$ and $\sum_i x_i \rho_z = 0$, while $\sum_i z_i \rho_z \neq 0$, where $(x,y,z)_i$ is the $(x,y,z)$-coordinate of the location of the charge.
As such, one can say the membrane operator is protected by the subsystem symmetries.

\subsubsection{Ground state degeneracy}
The ground state degeneracy of the quantum breathing pyrochlore model can be obtained by using the membrane operators.
The membrane operators allow tunneling between the various ground states, and as such we can generate a particular ground state from a given ground state by applying membrane operators.
We tabulate the numerically computed ground state degeneracy in Table ~\ref{tab:gsd} for different finite size cluster specified by $(L_{x},L_{y},L_{z})$.
We note that there are two procedures of determining the states satisfying the Gauss' law constraint: (i) a ``naive'' methodology where one enumerates over all the possible states to find those that satisfy Eq.~\ref{eq:gausscon}, and (ii) employment of the aforementioned membrane operators, which allow tunneling between the various states in the ground state manifold. 
We have explicitly verified that these methods agree for cases $L_{x}L_{y}L_{z}\leq 4$; for ease of numerical computation, it is advantageous to implement the membrane operator approach.

The computed ground state degeneracies shed a remarkable insight into the non-trivial nature of the quantum ground states.
Unlike the X-cube model \cite{PhysRevB.94.235157}, where the (logarithm of the) ground state degeneracy only depends on the perimeter, $L_{x}+L_{y}+L_{z}$ \cite{PhysRevB.94.235157,PhysRevB.95.245126,PhysRevB.96.195139,vijay2017isotropic,PhysRevX.8.031051}, in our model, the ground state degeneracy can differ even with the same perimeter or same volume. It is more complicated and depends on the lattice geometry.

We classify the cases as follows: (i) $L_{i}\geq2$ and $L_{j}=L_{k}=1$, (ii) $L_{i},L_{j}\geq2$ and $L_{k}=1$, and (iii) $L_{i}\geq2$ for all $i=x,y,z$.
In the case (i), the ground state degeneracy monotonically increases as a function of the length of the system, $L_{i}$.
In the cases (ii) and (iii), the ground state degeneracy does not monotonically increase as volume and perimeter increase, as mentioned previously. 
In each case of (ii) and (iii), the configuration having a larger perimeter has larger ground state degeneracy than configuration having smaller perimeter, regardless of the volume. %
This is because the number of times the membrane operators can be applied on the system depends on the perimeter; we recall that since we construct the ground states by applying the membrane operators, the ground state degeneracy is thus dependent on how many times we can apply the membrane operators. %
Since the membrane operators act on the planes in the system, the number of times they may be applied thus also depends on the number of planes in the system. 
The face-centered cubic geometry formed by the A-tetrahedra has $2L_{i}$ number of planes in each $i$-direction; 
in each direction, $L_{i}$ number of planes consists of A-tetrahedra sites on the vertex of the cube, and remaining $L_{i}$ number of planes consist of A-tetrahedron sites on the centre of the faces of the cube (Fig.~\ref{fig:FCCunit}). 
As such, the total number of the planes in a given geometry is $2(L_{x}+L_{y}+L_{z})$, i.e., double the perimeter. This subsequently implies that the number of the operation depends on the perimeter.
Therefore, having large perimeter leads to the large number of the ground state degeneracy. %
Furthermore, we find that the ground state degeneracy monotonically decreases as the volume increases for a fixed perimeter. 
This is because the number of independent constraints for the ground states (which are Gauss' law constraints) depends on the volume. 
The numbers of independent Gauss' law constraints is found to be $2L_{x}L_{y}L_{z}-1$ for case (i), and $4L_{x}L_{y}L_{z}$ for cases of (ii) and (iii), respectively. 
The reason why the number of independent Gauss' law constraints of case (i) is less than cases (ii) and (iii) is due to the finite size effect arising from the periodic boundary condition. 
We recall that we can find four units of Gauss' law constraints per one unit cell of FCC lattices (Fig.~\ref{fig:FCCunit}), but due to the periodic boundary conditions, case (i) has two independent units of Gauss' law constraints in one unit cell. As such, when they have the same perimeter, if we have case (ii) or (iii) rather than case (i), if we have a larger volume, then we have a smaller ground state degeneracy because there are a large number of independent constraints.

As a result, the ground state degeneracy of the system is non-extensive with volume and depends on the geometry of the system. 
This tendency is similar to previous fractonic phase of matters which show sub-extensive ground state degeneracy \cite{PhysRevA.83.042330,PhysRevB.95.245126,PhysRevB.94.235157,PhysRevB.95.245126,PhysRevB.96.195139,doi:10.1146/annurev-conmatphys-031218-013604,doi:10.1142/S0217751X20300033}. This suggests that the non-extensive behavior of the ground state degeneracy of our system is an indication that our system is indeed fractonic.

We now return back to the `possible' diagonal-$\mathbb{E}_{i,zz} \mathbb{E}_{j,zz}$ terms in $H'$; we recall that our above analysis focussed on `non-diagonal' perturbative terms in order to understand the delicacy of the ground state manifold and the possibility to tunnel between the multitude of states. 
Since such terms do not change the $E_{zz}$ quantum number, they do not introduce charged excitations nor do they change the structure of the described boundary charges.
To understand the role of such terms, we draw a comparison to quantum spin ice, where a [111] magnetic field does not destabilize the quantum spin ice phase for small field strengths (i.e. the quantum spin ice remains self-consistently stable) \cite{PhysRevLett.119.227204}.
It is certainly possible that the above diagonal terms may play a similar role and leave the fractonic phase stable for particular parameter choices. 
Indeed, understanding the ultimate role of this delicate interplay between all the perturbative terms would require extensive numerical simulations that we reserve for future work.
We can, nonetheless, glean the effects of such a diagonal term by simply considering the possible energy shift it may have on the ground-state manifold within first-order perturbation theory.
As we demonstrate in Appendix \ref{app_diagonal_terms_perturb}, in the thermodynamic limit, the ground state degeneracy though reduced, still remains dependent on lattice geometry, exponentially grows with system size, and is sub-extensive in system volume, all of which suggest the retention of fractonic properties.
For example, the remnant ground state degeneracy for $K>0$ (within first-order perturbation theory) is $\text{GSD}_{K>0}=2^{2L_{z}}$.
We nevertheless note that in an experimentally realizable finite-temperature setting, where the temperature is comparable to or larger than this split gap, such a splitting of the ground degeneracy would manifest as a quasi-degeneracy where the conclusions established above in Table \ref{tab:gsd} hold.

\begin{table}
\begin{tabular}{>{$}c<{$}|>{$}c<{$}|>{$}c<{$}|>{$}c<{$}|>{$}c<{$}|>{$}r<{$}|>{$}c<{$}}
\hline
L_{x}&L_{y}&L_{z}&\text{volume}&\text{perimeter}&\text{GSD}\;\;\;&\text{constraints}\\
\hline\hline
1&1&1&1&3&85&1\\
\hline
2&1&1&2&4&           1,333&3\\
3&1&1&3&5&         25,405&5\\
4&1&1&4&6&       535,333&7\\
5&1&1&5&7&  11,982,925&9\\
6&1&1&6&8&278,766,133&11\\
\hline
2&2&1&4  &5&       10,213&16\\
3&2&1&6  &6&     116,653&24\\
4&2&1&8  &7&  1,664,533&32\\
3&3&1&9  &7&     889,525&36\\
5&2&1&10&8&27,510,973&40\\
4&3&1&12&8&  9,103,453&48\\
\hline
2&2&2&8  &6&       49,541&32\\
3&2&2&12&7&     392,365&48\\
4&2&2&16&8&  4,201,589&64\\
3&3&2&18&8&  2,258,486&72\\
5&2&2&20&9&55,306,813&  80\\
4&3&2&24&9&18,470,173&  96\\
3&3&3&27&9&  9,912,253&108\\
\end{tabular}\caption{Table for the ground state degeneracy constructed by applying the membrane operators in terms of $(L_{x},L_{y},L_{z})$. 
The first block is the ground state degeneracy for $L_{x}=L_{y}=L_{z}=1$,
the second block is the ground state degeneracy for $L_{x}=L_{y}=L_{z}=1$, the third block is for $L_{i}\geq2$ and $L_{j}=L_{k}=1$, the last block is for $L_{i},L_{j}\geq2$ and $L_{k}=1$, and the last block is for $L_{i}\geq2$ for all $i=x,y,z$. %
If the volume, $L_{x}L_{y}L_{z}$, and the perimeter, $L_{x}+L_{y}+L_{z}$, are the same, then we have the same number of the ground state degeneracy. In the same block, if the perimeter is large and the volume is small, we may have a large number of the ground state degeneracy.
}\label{tab:gsd}
\end{table}
\subsection{Absence of magnetic field from finite-order of perturbation theory}\label{sec:absence_mag}

The membrane operators require the system size-dependent order of operation to return back to the ground state manifold (with the appropriate boundary conditions).
In the thermodynamic limit, the coefficient for such a perturbative process, or the membrane operator, is proportional to $(t/a_{\text{B},\textsf{T}_{1+}})^{L^{2}}$ which is system size-dependent where $t\ll a_{\text{B},\textsf{T}_{1+}}$ is the coefficient of the perturbation Hamiltonian containing the raising and lowering operators. 
Drawing an analogy with the Hamiltonian for the electromagnetism, $H=\tfrac{\epsilon}{2} E^{2}+\frac{1}{2\mu}B^{2}$, since the membrane operator generates the terms corresponds to $B^{2}$, we may regard the corresponding permeability ${1}/{\mu}\propto (t/a_{\text{B},\textsf{T}_{1+}})^{L^{2}}$ i.e. $1/\mu\rightarrow0$ in large system size.
This suggests that the corresponding ``speed of light'' is similiarly suppressed to zero, as $c \sim 1/\sqrt{\mu} \propto (t/a_{\text{B},\textsf{T}_{1+}})^{L^{2}/2}\rightarrow0$ in large system size.   %
This allows us to interpret that the photon in the breathing pyrochlore lattice as being extremely ``slow''.
Indeed, this interpretation is reminiscent of physics of quantum glassiness where the tunneling between two different ground states requires an exponentially long time, leading to glassy behaviour \cite{PhysRevLett.94.040402,doi:10.1080/14786435.2011.609152,PhysRevA.83.042330}. 
Analogously, we may regard our system as similarly requiring long time to tunnel between two different ground states, $t_{\text{char}}\sim t_{0}e^{(L^{2}/2)\ln(a_{\text{B},\textsf{T}_{1+}}/t)}$, where $t_0$ is a microscopic time scale.

The disappearance of the perturbative magnetic field term in the thermodynamic limit is in stark comparison to the magnetic field term that is generated at finite-order perturbation theory in quantum spin ice \cite{PhysRevB.69.064404,PhysRevB.94.075146} and previous higher-rank gauge theory constructions \cite{PhysRevB.95.115139,PhysRevB.96.035119,PhysRevB.74.224433,PhysRevD.81.104033,PhysRevB.97.235112}.
Indeed it is the complicated three-dimensional geometry of the breathing pyrochlore lattice that prohibits a finite-order perturbative process that allows tunnelling between the degenerate ground state manifold.
We recall that the application of a raising/lowering operator leads to charges being created in a three-dimensional volume as seen in Fig.~\ref{fig:charge_config}.
This is unlike the case of creating gauge charges along a line or a plane \cite{PhysRevB.95.115139,PhysRevB.96.035119,PhysRevB.74.224433,PhysRevD.81.104033,PhysRevB.97.235112}, where a perturbative pathway may be considered along a two-dimensional plane or a three-dimensional volume (respectively) that allows the charges to be ``wrapped around'' and eventually cancel each other.
We provide a simple example of such a process for gauge charges created in one and two-dimensions in Appendix~\ref{app_mag_field_none}.
In either case the ``corner charges'' (end of a line for one-dimensional line-charges or corners of a plane for two-dimensional plane-charges) are eliminated by appealing to a higher dimension than that of the charges; i.e. moving the one-dimensional line-charges around a two-dimensional plane, and two-dimensional planar-charges in an three-dimensional volume.
By extending the ideas of eliminating lower-dimensional charge configurations, it suggests that an additional (and not achievable in this setting) fourth dimension may be required to eliminate the corner charge on the breathing pyrochlore lattice.
We re-emphasize that the complicated geometry of the breathing pyrochlore lattice resulted in charges being created in three-dimensional volume space in the quantum model.
We note that even in these previous rank-2 U(1) models, this required higher-order perturbation processes in order to generate the magnetic field. For example, it required eighth-order in perturbation for the scenario of traceful magnetic fields, while in the traceless case a colossal thirty-second order of perturbation was needed \cite{PhysRevB.74.224433,PhysRevD.81.104033}.
We note that in previous higher-rank gauge theory constructions, it was that fact of having diagonal and off-diagonal electric field components reside on inequivalent lattice sites that allowed a finite-order perturbative process to connect the different ground states \cite{PhysRevB.74.224433,PhysRevD.81.104033,PhysRevB.95.115139,PhysRevB.96.035119,PhysRevB.97.235112}.
In our case, since diagonal and off-diagonal electric field components reside on the equivalent sites, lowering/raising operators of the electric field component lead to charges being created in a three-dimensional (tetragonal) volume regardless of whether it is a diagonal or off-diagonal component. As such, this suggests that we may not find such finite order perturbation processes, even if we have all the electric field components. %
Note that the only operator that allows tunneling between the ground states is the membrane operator. 
At the finite order perturbation theory in the thermodynamic limit, the afore-described diagonal perturbative terms lead to a small energy shift amongst the ground states.
It would be an interesting direction of future work to study the mixture of the diagonal terms with off-diagonal terms, and whether they may compete or cooperate with each other (with respect to how they may change the ground state).

\section{Discussion}\label{sec_discussions}

In this work, we provide a concrete model for fractonic quantum phases on the breathing pyrochlore lattice. 
In contrast to recently studied exactly-solvable fractonic models that involve interactions between a large number of particles/spins, the quantum model we consider involves bilinear interactions between spin-1/2 moments residing on the vertices of the corner-sharing tetrahedra. 
As such, this provides a more natural and realistic setting to realize such exotic quantum phases of matter. 

Though the previously studied classical model on the breathing pyrochlore lattice is captured within the framework of a rank-2 vector gauge theory \cite{PhysRevLett.124.127203}, we find that the quantum model has some sharp distinctions.
In particular, the electric field components do not commute (and satisfy an SU(2) algebra), and the conserved charge degree of freedom is the z-component of the vector charge, $\rho_z$, with the remaining components completing the SU(2) spinor algebra, $[\rho_x, \rho_y] = i \rho_z$.
These corresponding elementary spinor excitations are created in a quartet in three-dimensional space such that attempting to move a single particle results in a ``burst'' of collective quartet of spinor charges.
Furthermore, the ground state is found to have a degeneracy that is non-extensive with volume, yet strongly dependent on the geometrical configuration.
The immobility of the excitations compounded with a non-extensive (yet geometry dependent) ground state degeneracy is highly indicate of a fractonic phase of matter \cite{PhysRevA.83.042330,PhysRevB.95.245126,PhysRevB.94.235157,PhysRevB.95.245126,PhysRevB.96.195139,doi:10.1146/annurev-conmatphys-031218-013604,doi:10.1142/S0217751X20300033}.
Intriguingly, the quantum model we consider also lacks a local magnetic field term that connects the various quantum states of the degenerate manifold at finite order in perturbation theory.
This salient feature heralds the demise of any propagating photonic excitation, and the birth of glassy dynamics \cite{PhysRevLett.94.040402,doi:10.1080/14786435.2011.609152,PhysRevA.83.042330}, which is in sharp contrast with conventional graviton excitations in rank-2 gauge theories \cite{PhysRevB.74.224433,PhysRevD.81.104033}.

In classical limit of rank-2 vector gauge theory, we have the conservations of the total vector charge, $\int \bm{\rho}=0$, and angular momentum, $\int \bm{x}\times\bm{\rho}=0$ \cite{PhysRevB.95.115139,PhysRevB.96.035119}, but
it is not immediately apparent that there is a restriction for movement of $z$-charge along $z$ direction. 
In fact, the $z$-charge can move along $z$ direction, as was demonstrated in a microscopic model of rank-2 vector gauge theory on a simple cubic lattice \cite{PhysRevB.95.115139}.
However, in our microscopic model, due to the complicated three-dimensional orientation of the breathing pyrochlore lattice, as shown in Fig.~\ref{fig:charge_config}, increasing or decreasing the electric fields on A-tetrahedron leads to charges being created in a three-dimensional volume. 
Therefore, there is no such hopping term that allows $z$-charge to solely move along the $z$ direction. This is from the distinction between the microscopic quantum model and the classical limit. 

The model we consider is in the limit of particular energy scales that allows us to consider solely the diagonal components of the electric field i.e. the focussing on the corresponding classical ``light'' $\textsf{A}_{2}, \textsf{E}, \textsf{T}_{1-}$ modes.
Indeed, relaxing this condition may allow the introduction of off-diagonal electric field components (namely $\textsf{T}_{1+},  \textsf{T}_{2}$ modes) into the quantum model.
The virtue of our consideration is a clean closure of the corresponding algebra. 
It would be intriguing to explore whether the above properties of the breathing pyrochlore model survive with this relaxation of the coupling constant values.

\begin{acknowledgements}
We thank Han Yan and Daniel Bulmash for helpful discussions. 
This work was supported by the NSERC of Canada and the Center for Quantum Materials at the University of Toronto. 

\end{acknowledgements}

\appendix

\section{Normal mode representation of the microscopic interactions between spins on the breathing pyrochlore lattice}\label{app:normalcoeff}

A generalized nearest neighbour spin model on the breathing pyrochlore lattice involving antiferromagnetic Heisenberg, bond-dependent Dzyaloshinskii-Moriya (DM), Kitaev and Gamma interactions, is of the form given in Eq.~\ref{eq:generalspinmodel},
\begin{widetext}
\begin{align}
H=&\sum_{\braket{ij}\in\text{A}}\Big[J_{\text{A}}\mathbf{S}_{i}\cdot\mathbf{S}_{j}+D_{\text{A}}\hat{\mathbf{d}}_{ij}\cdot(\mathbf{S}_{i}\times\mathbf{S}_{j})+K_{\text{A},ij}^{\alpha}S_{i}^{\alpha}S_{j}^{\alpha}+\Gamma_{\text{A},ij}^{\gamma\delta}(S_{i}^{\gamma}S_{j}^{\delta}+S_{i}^{\delta}S_{j}^{\gamma})+E_{\text{A},0}\Big]\notag\\
&+\sum_{\braket{ij}\in\text{B}}\Big[J_{\text{B}}\mathbf{S}_{i}\cdot\mathbf{S}_{j}+D_{\text{B}}\hat{\mathbf{d}}_{ij}\cdot(\mathbf{S}_{i}\times\mathbf{S}_{j})+K_{\text{B},ij}^{\alpha}S_{i}^{\alpha}S_{j}^{\alpha}+\Gamma_{\text{B},ij}^{\gamma\delta}(S_{i}^{\gamma}S_{j}^{\delta}+S_{i}^{\delta}S_{j}^{\gamma})+E_{\text{B},0}\Big] \label{eq:generalspinmodel}\\
=&\frac{1}{2}\sum_{\text{A},\Gamma}a_{\text{A},\Gamma}m_{\text{A},\Gamma}^{2}+\frac{1}{2}\sum_{\text{B},\Gamma}a_{\text{B},\Gamma}m_{\text{B},\Gamma}^{2} \label{eq_normal_app},
\end{align}
\end{widetext}
where $J_{(\text{A},\text{B})}$, $D_{(\text{A},\text{B})}$ are the interaction coefficients of the Heisenberg and DM interaction, and $E_{(\text{A},\text{B}),0}$ is the constant energy shift on A (B)-tetrahedron, respectively, and $\hat{\mathbf{d}}_{ij}$ are the bond-dependent vectors defined in Ref.~\cite{PhysRevLett.124.127203}. 
For clarity, we note that,
\begin{align}
K_{(\text{A},\text{B}),01}=&K_{(\text{A},\text{B}),23}=K_{(\text{A},\text{B})}(1,0,0),\\ 
K_{(\text{A},\text{B}),02}=&K_{(\text{A},\text{B}),13}=K_{(\text{A},\text{B})}(0,1,0),\\
K_{(\text{A},\text{B}),03}=&K_{(\text{A},\text{B}),12}=K_{(\text{A},\text{B})}(0,0,1),\\
\Gamma_{(\text{A},\text{B}),01}=&-\Gamma_{(\text{A},\text{B}),23}=\Gamma_{\text{A}}\left(\begin{matrix}
0&0&0\\
0&0&1\\
0&1&0\\
\end{matrix}\right),\\
\Gamma_{(\text{A},\text{B}),02}=&-\Gamma_{(\text{A},\text{B}),13}=\Gamma_{\text{A}}\left(\begin{matrix}
0&0&1\\
0&0&0\\
1&0&0\\
\end{matrix}\right),\\
\Gamma_{(\text{A},\text{B}),03}=&-\Gamma_{(\text{A},\text{B}),12}=\Gamma_{\text{A}}\left(\begin{matrix}
0&1&0\\
1&0&0\\
0&0&0\\
\end{matrix}\right),
\end{align}
with $K_{(\text{A},\text{B})}$ and $\Gamma_{(\text{A},\text{B})}$ denoting the interaction coefficients of bond-dependent Kitaev and Gamma interactions, respectively.

The interacting Hamiltonian can be recast into a normal mode representation, as given in Eq.~\ref{eq_normal_app}.
The corresponding normal mode interaction coefficients are related to the microscopic interaction parameters via \cite{McClarty_2009} (dropped the A- and B- tetrahedron labels for brevity),
\begin{align}
a_{\textsf{A}_{2}}=&\;\frac{2E_{0}}{3}-J-\frac{4D}{\sqrt{2}}+K-4\Gamma,\\
a_{\textsf{E}}=&\;\frac{2E_{0}}{3}-J+\frac{2D}{\sqrt{2}}+K+2\Gamma,\\
a_{\textsf{T}_{1-}}=&\;\frac{2E_{0}}{3}-J+\frac{2D}{\sqrt{2}}-K-2\Gamma,\\
a_{\textsf{T}_{2}}=&\;\frac{2E_{0}}{3}-J-\frac{2D}{\sqrt{2}}-K+2\Gamma,\\
a_{\textsf{T}_{1+}}=&\;\frac{2E_{0}}{3}+3J+K.
\end{align}
We note that if $J_{(\text{A},\text{B})}$ is positive and larger than the other coefficients (i.e. $a_{(\text{A},\text{B}),\textsf{T}_{1+}} > 0$ is the largest coefficient), then we can take $\mathbf{m}_{(\text{A},\text{B}),\textsf{T}_{1+}}$=0. %

The generic interacting spin model in Eq.~\ref{eq:generalspinmodel} reduces to the microscopic spin model considered in Ref.~\cite{PhysRevLett.124.127203} by setting $K$, $\Gamma$ and $E_{0}$ to zero.
In particular, the coefficients reduce to,
\begin{align}
a_{\textsf{A}_{2}}=&-J-\frac{4D}{\sqrt{2}},\\
a_{\textsf{E}}=&\;a_{\textsf{T}_{1-}}=-J+\frac{2D}{\sqrt{2}},\\
a_{\textsf{T}_{2}}=&-J-\frac{2D}{\sqrt{2}}\\
a_{\textsf{T}_{1+}}=&\;3J.
\end{align}

In the main text, we consider $a_{\text{A},\textsf{A}_{2}}=a_{\text{A},\textsf{E}}$ and the hierarchy, $a_{\text{A},\textsf{A}_{2}}=a_{\text{A},\textsf{E}}<a_{\text{A},\textsf{T}_{1-}}<a_{\text{A},\textsf{T}_{2}}<a_{\text{A},\text{T}_{1+}}$, which give us the light normal modes $m_{\text{A},\textsf{A}_{2}}$ and $\mathbf{m}_{\text{A},\textsf{E}}$, and heavy normal modes, $\mathbf{m}_{\text{A},\textsf{T}_{1\pm}}$ and $m_{\text{A},\textsf{T}_{2}}$.
These conditions can be easily achieved from the microscopic interactions of the generic spin model.
For instance, for $a_{\text{A},\textsf{A}_{2}}=a_{\text{A},\textsf{E}}$, one can take $\Gamma_{\text{A}}=-D_{\text{A}}/\sqrt{2}$. And, to satisfy the aforementioned hierarchy of energies, one can take $D_{\text{A}}=-|D_{\text{A}}|<0$, $K_{\text{A}}=-|K_{\text{A}}| < -\sqrt{2}|D_{\text{A}}|<0$, $\Gamma_{\text{A}}=|D_{\text{A}}|/\sqrt{2}>0$, and $J_{\text{A}}>(\sqrt{2}|D_{\text{A}}|+|K_{\text{A}}|)/2>0$.
These lead to,
\begin{align}
a_{\text{A},\textsf{A}_{2}}=&\;a_{\text{A},\textsf{E}}=-J_{\text{A}}-|K_{\text{A}}|,\\
a_{\text{A},\textsf{T}_{1-}}=&-J_{\text{A}}-\frac{4|D_{\text{A}}|}{\sqrt{2}}+|K_{\text{A}}|,\\
a_{\text{A},\textsf{T}_{2}}=&-J_{\text{A}}+\frac{4|D_{\text{A}}|}{\sqrt{2}}+|K_{\text{A}}|,\\
a_{\text{A},\textsf{T}_{1+}}=&\;3J_{\text{A}}-|K_{\text{A}}|,
\end{align}
where we also set $E_{0}=0$, to thus recover the aforementioned hierarchy of energies.

\begin{table*}[t]
\begin{tabular}{|>{$}c<{$}|>{$}c<{$}|}
\hline
\text{normal mode}&\text{definition}\\
\hline
m_{\textsf{A}_{2}}&\tfrac{1}{2\sqrt{3}}[(S_{0}^{x}+S_{0}^{y}+S_{0}^{z})+(S_{1}^{x}-S_{1}^{y}-S_{1}^{z})+(-S_{2}^{x}+S_{2}^{y}-S_{2}^{z})+(-S_{3}^{x}-S_{3}^{y}+S_{3}^{z})]\\
\hline
\mathbf{m}_{\textsf{E}}&\left( 
\begin{matrix}\tfrac{1}{2\sqrt{6}}[(-2S_{0}^{x}+S_{0}^{y}+S_{0}^{z})+(-2S_{1}^{x}-S_{1}^{y}-S_{1}^{y})+(2S_{2}^{x}+S_{2}^{y}-S_{2}^{z})+(2S_{3}^{x}-S_{3}^{y}+S_{3}^{z})] \\
\tfrac{1}{2\sqrt{2}}[(-S_{0}^{y}+S_{0}^{z})+(S_{1}^{y}-S_{1}^{z})+(-S_{2}^{y}-S_{2}^{z})+(S_{3}^{y}+S_{3}^{z})]
\end{matrix}\right)\\
\hline
\mathbf{m}_{\textsf{T}_{2}}&\left(\begin{matrix}
\tfrac{1}{2\sqrt{2}}[(-S_{0}^{y}+S_{0}^{z})+(S_{1}^{y}-S_{1}^{z})+(S_{2}^{y}+S_{2}^{z})+(-S_{3}^{y}-S_{3}^{z})]\\
\tfrac{1}{2\sqrt{2}}[(S_{0}^{x}-S_{0}^{z})+(-S_{1}^{x}-S_{1}^{z})+(-S_{2}^{x}+S_{2}^{z})+(S_{3}^{x}+S_{3}^{z})]\\
\tfrac{1}{2\sqrt{2}}[(-S_{0}^{x}+S_{0}^{y})+(S_{1}^{x}+S_{1}^{y})+(-S_{2}^{x}-S_{2}^{y})+(S_{3}^{x}-S_{3}^{y})]
\end{matrix}\right)\\\hline
\mathbf{m}_{\textsf{T}_{1+}}&\left(\begin{matrix}
\tfrac{1}{2}[S_{0}^{x}+S_{1}^{x}+S_{2}^{x}+S_{3}^{x}]\\
\tfrac{1}{2}[S_{0}^{y}+S_{1}^{y}+S_{2}^{y}+S_{3}^{y}]\\
\tfrac{1}{2}[S_{0}^{z}+S_{1}^{z}+S_{2}^{z}+S_{3}^{z}]
\end{matrix}\right)\\\hline
\mathbf{m}_{\textsf{T}_{1-}}&\left(\begin{matrix}
\tfrac{-1}{2\sqrt{2}}[(S_{0}^{y}+S_{0}^{z})+(-S_{1}^{y}-S_{1}^{z})+(-S_{2}^{y}+S_{2}^{z})+(S_{3}^{y}-S_{3}^{z})]\\
\tfrac{-1}{2\sqrt{2}}[(S_{0}^{x}+S_{0}^{z})+(-S_{1}^{x}+S_{1}^{z})+(-S_{2}^{x}-S_{2}^{z})+(S_{3}^{x}-S_{3}^{z})]\\
\tfrac{-1}{2\sqrt{2}}[(S_{0}^{x}+S_{0}^{y})+(-S_{1}^{x}+S_{1}^{y})+(S_{2}^{x}-S_{2}^{y})+(-S_{3}^{x}-S_{3}^{y})]
\end{matrix}\right)\\\hline
\end{tabular}\caption{The definition of the normal modes on the A- and B-tetrahedra in terms of the spin degrees of freedom.}\label{tab:def_normal}
\end{table*}

\section{Derivation of Gauss' laws}\label{app:gausscontinu}

The classical Gauss's law constraint arises from taking the $\mathbf{m}_{(\text{A},\text{B}), \textsf{T}_{1+}} = 0$.
Using Table \ref{tab:def_normal}, the B-normal mode can be rewritten in terms of the normal modes of the surrounding four A-tetrahedron surrounding a given B-tetrahedron,
\begin{widetext}
\begin{align}
m_{\text{B},\textsf{T}_{1+}}^{x}=&\frac{1}{4}\sum_{\alpha=0}^{3}[m_{\alpha,\textsf{T}_{1+}}^{x}+\frac{c_{z,\alpha}}{\sqrt{2}}(m_{\alpha,\textsf{T}_{1-}}^{y}-m_{\alpha,\textsf{T}_{2}}^{y})+\frac{c_{y,\alpha}}{\sqrt{2}}(m_{\alpha,\textsf{T}_{1-}}^{z}+m_{\alpha,\textsf{T}_{2}}^{z})-c_{x,\alpha}(\tfrac{1}{\sqrt{3}}m_{\alpha,\textsf{A}_{2}}-\sqrt{\tfrac{2}{3}}m_{\alpha,\textsf{E}}^{1})
]\notag\\
\approx&m_{\text{A},\textsf{T}_{1+}}^{x}(0)+\frac{a_{d}}{4\sqrt{2}}\partial_{z}(m_{\text{A},\textsf{T}_{1-}}^{y}-m_{\text{A},\textsf{T}_{2}}^{y})+\frac{a_{d}}{4\sqrt{2}}\partial_{y}(m_{\text{A},\textsf{T}_{1-}}^{z}+m_{\text{A},\textsf{T}_{2}}^{z})-\frac{a_{d}}{4}\partial_{x}(\tfrac{1}{\sqrt{3}}m_{\text{A},\textsf{A}_{2}}-\sqrt{\tfrac{2}{3}}m_{\text{A},\textsf{E}}^{1}), \label{eq:mbtp1}\\
m_{\text{B},\textsf{T}_{1+}}^{y}=&\frac{1}{4}\sum_{\alpha=0}^{3}[m_{\alpha,\textsf{T}_{1+}}^{y}+\frac{c_{x,\alpha}}{\sqrt{2}}(m_{\alpha,\textsf{T}_{1-}}^{z}-m_{\alpha,\textsf{T}_{2}}^{z})+\frac{c_{z,\alpha}}{\sqrt{2}}(m_{\alpha,\textsf{T}_{1-}}^{x}+m_{\alpha,\textsf{T}_{2}}^{x})-c_{y,\alpha}(\tfrac{1}{\sqrt{3}}m_{\alpha,\textsf{A}_{2}}+\tfrac{1}{\sqrt{6}}m_{\alpha,\textsf{E}}^{1}-\tfrac{1}{\sqrt{2}}m_{\alpha,\textsf{E}}^{2})
]\notag\\
\approx&m_{\text{A},\textsf{T}_{1+}}^{y}(0)+\frac{a_{d}}{4\sqrt{2}}\partial_{x}(m_{\text{A},\textsf{T}_{1-}}^{z}-m_{\text{A},\textsf{T}_{2}}^{z})+\frac{a_{d}}{4\sqrt{2}}\partial_{z}(m_{\text{A},\textsf{T}_{1-}}^{x}+m_{\text{A},\textsf{T}_{2}}^{x})-\frac{a_{d}}{4}\partial_{y}(\tfrac{1}{\sqrt{3}}m_{\text{A},\textsf{A}_{2}}+\tfrac{1}{\sqrt{6}}m_{\text{A},\textsf{E}}^{1}-\tfrac{1}{\sqrt{2}}m_{\text{A},\textsf{E}}^{2}),\label{eq:mbtp2}\\
m_{\text{B},\textsf{T}_{1+}}^{z}=&\frac{1}{4}\sum_{\alpha=0}^{3}[m_{\alpha,\textsf{T}_{1+}}^{z}+\frac{c_{y,\alpha}}{\sqrt{2}}(m_{\alpha,\textsf{T}_{1-}}^{x}-m_{\alpha,\textsf{T}_{2}}^{x})+\frac{c_{x,\alpha}}{\sqrt{2}}(m_{\alpha,\textsf{T}_{1-}}^{y}+m_{\alpha,\textsf{T}_{2}}^{y})-c_{z,\alpha}(\tfrac{1}{\sqrt{3}}m_{\alpha,\textsf{A}_{2}}+\tfrac{1}{\sqrt{6}}m_{\alpha,\textsf{E}}^{1}+\tfrac{1}{\sqrt{2}}m_{\alpha,\textsf{E}}^{2})
]\notag\\
\approx&m_{\text{A},\textsf{T}_{1+}}^{z}(0)+\frac{a_{d}}{4\sqrt{2}}\partial_{y}(m_{\text{A},\textsf{T}_{1-}}^{x}-m_{\text{A},\textsf{T}_{2}}^{x})+\frac{a_{d}}{4\sqrt{2}}\partial_{x}(m_{\text{A},\textsf{T}_{1-}}^{y}+m_{\text{A},\textsf{T}_{2}}^{y})-\frac{a_{d}}{4}\partial_{z}(\tfrac{1}{\sqrt{3}}m_{\text{A},\textsf{A}_{2}}+\tfrac{1}{\sqrt{6}}m_{\text{A},\textsf{E}}^{1}+\tfrac{1}{\sqrt{2}}m_{\text{A},\textsf{E}}^{2}),\label{eq:mbtp3}
\end{align}
where on the right side of the equalities the subscript $\alpha=0,1,2,3$ indicates the A-tetrahedron sharing a $0,1,2,3$ site on the B-tetrahedron (Fig.~\ref{fig:convention} in the main text) so we summed over the sublattice sites, 
 $c_{x,\alpha}=(-1,-1,1,1)$, $c_{y,\alpha}=(-1,1,-1,1)$, $c_{z,\alpha}=(-1,1,1,-1)$, $a_{d}$ is the lattice spacing constant, and $m_{\text{A},\textsf{T}_{1+}}^{i}(0)$ means $m_{\text{A},\textsf{T}_{1+}}^{i}$ at the origin of the gradient expansion. In Eq.~\ref{eq:mbtp1}, \ref{eq:mbtp2}, and \ref{eq:mbtp3}, we take a continuum limit.
In the continuum limit, if we take $\mathbf{m}_{\text{(A,B)},\textsf{T}_{1+}}=0$ and multiply them by $4\sqrt{2}$, we have
\begin{align}
&-\partial_{x}(\sqrt{\tfrac{2}{3}}m_{\text{A},\textsf{A}_{2}}-\tfrac{2}{\sqrt{3}}m_{\text{A},\textsf{E}}^{1})+\partial_{y}(m_{\text{A},\textsf{T}_{1-}}^{z}+m_{\text{A},\textsf{T}_{2}}^{z})+\partial_{z}(m_{\text{A},\textsf{T}_{1-}}^{y}-m_{\text{A},\textsf{T}_{2}}^{y})=0,\\
&\partial_{x}(m_{\text{A},\textsf{T}_{1-}}^{z}+m_{\text{A},\textsf{T}_{2}}^{z})-\partial_{y}(\sqrt{\tfrac{2}{3}}m_{\text{A},\textsf{A}_{2}}+\tfrac{1}{\sqrt{3}}m_{\text{A},\textsf{E}}^{1}-m_{\text{A},\textsf{E}}^{2})+\partial_{z}(m_{\text{A},\textsf{T}_{1-}}^{x}-m_{\text{A},\textsf{T}_{2}}^{x})=0,\\
&\partial_{x}(m_{\text{A},\textsf{T}_{1-}}^{y}+m_{\text{A},\textsf{T}_{2}}^{y})+\partial_{y}(m_{\text{A},\textsf{T}_{1-}}^{x}-m_{\text{A},\textsf{T}_{2}}^{x})-\partial_{z}(\sqrt{\tfrac{2}{3}}m_{\text{A},\textsf{A}_{2}}+\tfrac{1}{\sqrt{3}}m_{\text{A},\textsf{E}}^{1}+m_{\text{A},\textsf{E}}^{2})=0,
\end{align}
and we can rewrite them as follows:
\begin{align}
&
\tfrac{2}{\sqrt{3}}\left(\begin{matrix}
\partial_{x}m_{\text{A},\textsf{E}}^{1}\\
-\tfrac{1}{2}\partial_{y}m_{\text{A},\textsf{E}}^{1}+\tfrac{\sqrt{3}}{2}\partial_{y}m_{\text{A},\textsf{E}}^{2}\\
-\tfrac{1}{2}\partial_{z}m_{\text{A},\textsf{E}}^{1}-\tfrac{\sqrt{3}}{2}\partial_{z}m_{\text{A},\textsf{E}}^{2}\\
\end{matrix}\right)+
\left(\begin{matrix}
\partial_{y}m_{\text{A},\textsf{T}_{1-}}^{z}+\partial_{z}m_{\text{A},\textsf{T}_{1-}}^{y}\\
\partial_{z}m_{\text{A},\textsf{T}_{1-}}^{x}+\partial_{x}m_{\text{A},\textsf{T}_{1-}}^{z}\\
\partial_{x}m_{\text{A},\textsf{T}_{1-}}^{y}+\partial_{y}m_{\text{A},\textsf{T}_{1-}}^{x}
\end{matrix}\right)
-\sqrt{\tfrac{2}{3}}\nabla m_{\text{A},\textsf{A}_{2}}-\nabla\times \mathbf{m}_{\text{A},\textsf{T}_{2}}\\
&=\nabla\cdot(\mathbf{E}_{\text{A}}^{\text{trace}}+\mathbf{E}_{\text{A}}^{\text{sym}}+\mathbf{E}_{\text{A}}^{\text{antisym}})=0
\end{align}
where
\begin{align}
(\mathbf{E}_{\text{A}}^{\text{trace}})_{ij}=&-\sqrt{\tfrac{2}{3}}m_{\text{A},\textsf{A}_{2}}\delta_{ij}, &
(\mathbf{E}_{\text{A}}^{\text{antisym}})_{ij}=&-\epsilon_{ijk}m_{\text{A},\textsf{T}_{2}}^{k},&
\mathbf{E}_{\text{A}}^{\text{sym}}=&%
\left(\begin{matrix}
\tfrac{2}{\sqrt{3}}m_{\text{A},\textsf{E}}^{1} & m_{\text{A},\textsf{T}_{1-}}^{z} & m_{\text{A},\textsf{T}_{1-}}^{y}\\
m_{\text{A},\textsf{T}_{1-}}^{z} & -\tfrac{1}{\sqrt{3}}m_{\text{A},\textsf{E}}^{1}+m_{\text{A},\textsf{E}}^{2} & m_{\text{A},\textsf{T}_{1-}}^{x}\\
m_{\text{A},\textsf{T}_{1-}}^{y} & m_{\text{A},\textsf{T}_{1-}}^{x} & -\tfrac{1}{\sqrt{3}}m_{\text{A},\textsf{E}}^{1}-m_{\text{A},\textsf{E}}^{2}\\
\end{matrix}\right).
\end{align}
The traceful electric fields defined in main text are defined as
\begin{align}
\mathbb{E}_{\text{A},xx}=&\tfrac{2}{\sqrt{3}}m_{\text{A},\textsf{A}_{2}}-2\sqrt{\tfrac{2}{3}}m_{\text{A},\textsf{E}}^{1},&
\mathbb{E}_{\text{A},yy}=&\tfrac{2}{\sqrt{3}}m_{\text{A},\textsf{A}_{2}}+\sqrt{\tfrac{2}{3}}m_{\text{A},\textsf{E}}^{1}-\sqrt{2}m_{\text{A},\textsf{E}}^{2},&
\mathbb{E}_{\text{A},zz}=&\tfrac{2}{\sqrt{3}}m_{\text{A},\textsf{A}_{2}}+\sqrt{\tfrac{2}{3}}m_{\text{A},\textsf{E}}^{1}+\sqrt{2}m_{\text{A},\textsf{E}}^{2},&\\
\mathbb{E}_{\text{A},xy}=&-\sqrt{2}m_{\text{A},\textsf{T}_{1-}}^{z},&
\mathbb{E}_{\text{A},yz}=&-\sqrt{2}m_{\text{A},\textsf{T}_{1-}}^{x},&
\mathbb{E}_{\text{A},zx}=&-\sqrt{2}m_{\text{A},\textsf{T}_{1-}}^{y}.&
\end{align}
Here, the diagonal components satisfy the SU(2) algebra,
\begin{align}
[\mathbb{E}_{\text{A},i}\mathbb{E}_{\text{A}',j}]=i\delta_{\text{A}\text{A}'}\epsilon_{ijk}\mathbb{E}_{\text{A},k}\quad\{i,j,k\}\in\{xx,yy,zz\}.
\end{align}
 because $\mathbb{E}_{\text{A},i}=-\sum_{\alpha=0}^{3}c_{i,a}S_{a}^{i}$ where $a$ is the site index of spins on A-tetrahedron.

\section{Charge-neutral quantum ground state configurations}
\label{app_charge_less_states}
In this section, we show 85 charge-neutral configurations for the ground states which satisfy $(a-b-c+d)=0$ (Eq.~\ref{eq:gausscon}) for $(L_{x},L_{y},L_{z})=(1,1,1)$ where $a,b,c,d\in\{\pm2,\pm1,0\}$ are the quantum numbers of the electric fields on each A-tetrahedron. The result is shown in Table.~\ref{tab:grnd111}. 

\begin{table}[h]
\begin{tabular}{|>{$}c<{$}|>{$}c<{$}|>{$}c<{$}|>{$}c<{$}||>{$}c<{$}|>{$}c<{$}|>{$}c<{$}|>{$}c<{$}||>{$}c<{$}|>{$}c<{$}|>{$}c<{$}|>{$}c<{$}|}\hline
a&b&c&d &a&b&c&d &a&b&c&d \\
\hline\hline
\pm2&\pm2&\pm2&\pm2 &
\pm2&\mp2&\pm2&\mp2 &
\pm1&\mp2&\pm2&0 
\\
\pm2&\pm2&\pm1&\pm1 &
\pm1&\pm2&\pm1&\pm2 &
\pm1&\mp1&\pm1&\mp1 
\\
\pm2&\pm2&0&0 &
\pm1&\pm2&0&\pm1 &
\pm1&\mp1&0&\mp2 
\\
\pm2&\pm2&\mp1&\mp1 &
\pm1&\pm2&\mp1&0 &
\pm1&\mp2&\pm2&\mp1 
\\
\pm2&\pm2&\mp2&\mp2 &
\pm1&\pm2&\mp2&\mp1 &
\pm1&\mp2&\pm1&\mp2 
\\
\pm2&\pm1&\pm2&\pm1 &
\pm1&\pm1&\pm2&\pm2 &
0&\pm2&0&\pm2 
\\
\pm2&\pm1&\pm1&0 &
\pm1&\pm1&\pm1&\pm1 &
0&\pm2&\mp1&\pm1 
\\
\pm2&\pm1&0&\mp1 &
\pm1&\pm1&0&0 &
0&\pm2&\mp2&0 
\\
\pm2&\pm1&\mp1&\mp1 &
\pm1&\pm1&\mp1&\mp1 &
0&\pm1&\pm1&\pm2 
\\
\pm2&0&\pm2&0 &
\pm1&\pm1&\mp2&\mp2 &
0&\pm1&0&\pm1 
\\
\pm2&0&\pm1&\mp1 &
\pm1&0&\pm2&\pm1 &
0&\pm1&\mp1&0 
\\
\pm2&0&0&\mp2 &
\pm1&0&\pm1&0 &
0&\pm1&\mp2&\mp1 
\\
\pm2&\mp1&\pm2&\mp1 &
\pm1&0&0&\mp1 &
0&0&\pm2&\pm2 
\\
\pm2&\mp1&\pm1&\mp2 &
\pm1&0&\mp1&\mp2 &
0&0&\pm1&\pm1 
\\
\cline{1-8}
\multicolumn{8}{c||}{}
&0&0&0&0 \\
\cline{9-12}
\end{tabular}
\caption{The 85 charge-neutral configurations of the ground states satisfying Eq.~\ref{eq:gausscon}, $(a-b-c+d)=0$. $a,b,c,d$ are the electric field quantum numbers with $a,b,c,d\in\{\pm2,\pm1,0\}$ on $0,1,2,3$-th A-tetrahedra (Fig.~\ref{fig:convention} in the main text), respectively.}\label{tab:grnd111}
\end{table}

\section{Normal modes on B-tetrahedron in terms of normal modes on A-tetrahedra}\label{app:BtoA}
The normal modes on B-tetrahedron in terms of normal modes on A-tetrahedra are as follows (where $c_{x,\alpha}=(-1,-1,1,1)$, $c_{y,\alpha}=(-1,1,-1,1)$, and $c_{z,\alpha}=(-1,1,1,-1)$):

\begin{align}
m_{\text{B},\textsf{A}_{2}}=&\frac{1}{4}\sum_{\alpha=0}^{3}[m_{\alpha,\textsf{A}_{2}}+\sqrt{\frac{2}{3}}(c_{x,\alpha}m_{\alpha,\textsf{T}_{1-}}^{x}+c_{y,\alpha}m_{\alpha,\textsf{T}_{1-}}^{y}+c_{z,\alpha}m_{\alpha,\textsf{T}_{1-}}^{z})-\frac{1}{\sqrt{3}}(c_{x,\alpha}m_{\alpha,\textsf{T}_{1+}}^{x}+c_{y,\alpha}m_{\alpha,\textsf{T}_{1+}}^{y}+c_{z,\alpha}m_{\alpha,\textsf{T}_{1+}}^{z})],\\
\notag m_{\text{B},\textsf{E}}^{1}=&\frac{1}{4}\sum_{\alpha=0}^{3}[m_{\alpha,\textsf{E}}^{1}+\frac{1}{2\sqrt{3}}(2c_{x,\alpha}m_{\alpha,\textsf{T}_{1-}}^{x}-c_{y,\alpha}m_{\alpha,\textsf{T}_{1-}}^{y}-c_{z,\alpha}m_{\alpha,\textsf{T}_{1-}}^{z})+\frac{1}{\sqrt{6}}(2c_{x,\alpha}m_{\alpha,\textsf{T}_{1+}}^{x}-c_{y,\alpha}m_{\alpha,\textsf{T}_{1+}}^{y}-c_{z,\alpha}m_{\alpha,\textsf{T}_{1+}}^{z})\\
&\quad\quad\quad+\frac{\sqrt{3}}{2}(c_{y,\alpha}m_{\alpha,\textsf{T}_{2}}^{y}-c_{z,\alpha}m_{\alpha,\textsf{T}_{2}}^{z})],\\
m_{\text{B},\textsf{E}}^{2}=&\frac{1}{4}\sum_{\alpha=0}^{3}[m_{\textsf{E}}^{2}+\frac{1}{2}(c_{y,\alpha}m_{\textsf{T}_{1-}}^{y}-c_{z,\alpha}m_{\textsf{T}_{1-}}^{z})+\frac{1}{\sqrt{2}}(m_{\textsf{T}_{1+}}^{y}-c_{z,\alpha}m_{\textsf{T}_{1+}}^{z})-(2c_{x,\alpha}m_{\textsf{T}_{2}}^{x}-c_{y,\alpha}m_{\textsf{T}_{2}}^{y}-c_{z,\alpha}m_{\textsf{T}_{2}}^{z})],\\
m_{\text{B},\textsf{T}_{1-}}^{x}
=&\frac{1}{4}\sum_{\alpha=0}^{3}[m_{\alpha,\textsf{T}_{1-}}^{x}-\frac{c_{z,\alpha}}{2}(m_{\alpha,\textsf{T}_{2}}^{y}+m_{\alpha,\textsf{T}_{1-}}^{y}-\sqrt{2}m_{\alpha,\textsf{T}_{1+}}^{y})+\frac{c_{y,\alpha}}{2}(m_{\alpha,\textsf{T}_{2}}^{z}-m_{\alpha,\textsf{T}_{1-}}^{z}+\sqrt{2}m_{\alpha,\textsf{T}_{1+}}^{z})\notag\\
&\quad\quad\quad+c_{x,\alpha}(\sqrt{\tfrac{2}{3}}m_{\alpha,\textsf{A}_{2}}+\tfrac{1}{\sqrt{3}}m_{\alpha,\textsf{E}}^{1})
],\\
m_{\text{B},\textsf{T}_{1-}}^{y}
=&\frac{1}{4}\sum_{\alpha=0}^{3}[m_{\alpha,\textsf{T}_{1-}}^{y}-\frac{c_{x,\alpha}}{2}(m_{\alpha,\textsf{T}_{2}}^{z}+m_{\alpha,\textsf{T}_{1-}}^{z}-\sqrt{2}m_{\alpha,\textsf{T}_{1+}}^{z})+\frac{c_{z,\alpha}}{2}(m_{\alpha,\textsf{T}_{2}}^{x}-m_{\alpha,\textsf{T}_{1-}}^{x}+\sqrt{2}m_{\alpha,\textsf{T}_{1+}}^{x})\notag\\
&\quad\quad\quad+c_{y,\alpha}(\sqrt{\tfrac{2}{3}}m_{\alpha,\textsf{A}_{2}}-\tfrac{1}{2\sqrt{3}}m_{\alpha,\textsf{E}}^{1}+\tfrac{1}{2}m_{\alpha,\textsf{E}}^{2})
]\\
m_{\text{B},\textsf{T}_{1-}}^{z}
=&\frac{1}{4}\sum_{\alpha=0}^{3}[m_{\alpha,\textsf{T}_{1-}}^{z}-\frac{c_{y,\alpha}}{2}(m_{\alpha,\textsf{T}_{2}}^{x}+m_{\alpha,\textsf{T}_{1-}}^{x}-\sqrt{2}m_{\alpha,\textsf{T}_{1+}}^{x})+\frac{c_{x,\alpha}}{2}(m_{\alpha,\textsf{T}_{2}}^{y}-m_{\alpha,\textsf{T}_{1-}}^{y}+\sqrt{2}m_{\alpha,\textsf{T}_{1+}}^{y})\notag\\
&\quad\quad\quad+c_{z,\alpha}(\sqrt{\tfrac{2}{3}}m_{\alpha,\textsf{A}_{2}}-\tfrac{1}{2\sqrt{3}}m_{\alpha,\textsf{E}}^{1}-\tfrac{1}{2}m_{\alpha,\textsf{E}}^{2})
],\\
m_{\text{B},\textsf{T}_{2}}^{x}=&\frac{1}{4}\sum_{\alpha=0}^{3}[m_{\alpha,\textsf{T}_{2}}^{x}+\frac{c_{z,\alpha}}{2}(m_{\alpha,\textsf{T}_{2}}^{y}+m_{\alpha,\textsf{T}_{1-}}^{y}+\sqrt{2}m_{\alpha,\textsf{T}_{1+}}^{y})+\frac{c_{y,\alpha}}{2}(m_{\alpha,\textsf{T}_{2}}^{z}-m_{\alpha,\textsf{T}_{1-}}^{z}-\sqrt{2}m_{\alpha,\textsf{T}_{1+}}^{z})\notag\\
&\quad\quad\quad-c_{x,\alpha}m_{\alpha,\textsf{E}}^{2}]\\
m_{\text{B},\textsf{T}_{2}}^{y}=&\frac{1}{4}\sum_{\alpha=0}^{3}[m_{\alpha,\textsf{T}_{2}}^{y}+\frac{c_{x,\alpha}}{2}(m_{\alpha,\textsf{T}_{2}}^{z}+m_{\alpha,\textsf{T}_{1-}}^{z}+\sqrt{2}m_{\alpha,\textsf{T}_{1+}}^{z})+\frac{c_{z,\alpha}}{2}(m_{\alpha,\textsf{T}_{2}}^{x}-m_{\alpha,\textsf{T}_{1-}}^{x}-\sqrt{2}m_{\alpha,\textsf{T}_{1+}}^{x})\notag\\
&\quad\quad\quad-c_{y,\alpha}(-\frac{\sqrt{3}}{2}m_{\alpha,\textsf{E}}^{1}-\frac{1}{2}m_{\alpha,\textsf{E}}^{2})]\\
m_{\text{B},\textsf{T}_{2}}^{z}=&\frac{1}{4}\sum_{\alpha=0}^{3}[m_{\alpha,\textsf{T}_{2}}^{z}+\frac{c_{y,\alpha}}{2}(m_{\alpha,\textsf{T}_{2}}^{x}+m_{\alpha,\textsf{T}_{1-}}^{x}+\sqrt{2}m_{\alpha,\textsf{T}_{1+}}^{x})+\frac{c_{x,\alpha}}{2}(m_{\alpha,\textsf{T}_{2}}^{y}-m_{\alpha,\textsf{T}_{1-}}^{y}-\sqrt{2}m_{\alpha,\textsf{T}_{1+}}^{y})\notag\\
&\quad\quad\quad-c_{z,\alpha}(\frac{\sqrt{3}}{2}m_{\alpha,\textsf{E}}^{1}-\frac{1}{2}m_{\alpha,\textsf{E}}^{2})]\\
m_{\text{B},\textsf{T}_{1+}}^{x}=&\frac{1}{4}\sum_{\alpha=0}^{3}[m_{\alpha,\textsf{T}_{1+}}^{x}+\frac{c_{z,\alpha}}{\sqrt{2}}(m_{\alpha,\textsf{T}_{1-}}^{y}-m_{\alpha,\textsf{T}_{2}}^{y})+\frac{c_{y,\alpha}}{\sqrt{2}}(m_{\alpha,\textsf{T}_{1-}}^{z}+m_{\alpha,\textsf{T}_{2}}^{z})-c_{x,\alpha}(\tfrac{1}{\sqrt{3}}m_{\alpha,\textsf{A}_{2}}-\sqrt{\tfrac{2}{3}}m_{\alpha,\textsf{E}}^{1})
]\\
m_{\text{B},\textsf{T}_{1+}}^{y}=&\frac{1}{4}\sum_{\alpha=0}^{3}[m_{\alpha,\textsf{T}_{1+}}^{y}+\frac{c_{x,\alpha}}{\sqrt{2}}(m_{\alpha,\textsf{T}_{1-}}^{z}-m_{\alpha,\textsf{T}_{2}}^{z})+\frac{c_{z,\alpha}}{\sqrt{2}}(m_{\alpha,\textsf{T}_{1-}}^{x}+m_{\alpha,\textsf{T}_{2}}^{x})-c_{y,\alpha}(\tfrac{1}{\sqrt{3}}m_{\alpha,\textsf{A}_{2}}+\tfrac{1}{\sqrt{6}}m_{\alpha,\textsf{E}}^{1}-\tfrac{1}{\sqrt{2}}m_{\alpha,\textsf{E}}^{2})
]\\
m_{\text{B},\textsf{T}_{1+}}^{z}=&\frac{1}{4}\sum_{\alpha=0}^{3}[m_{\alpha,\textsf{T}_{1+}}^{z}+\frac{c_{y,\alpha}}{\sqrt{2}}(m_{\alpha,\textsf{T}_{1-}}^{x}-m_{\alpha,\textsf{T}_{2}}^{x})+\frac{c_{x,\alpha}}{\sqrt{2}}(m_{\alpha,\textsf{T}_{1-}}^{y}+m_{\alpha,\textsf{T}_{2}}^{y})-c_{z,\alpha}(\tfrac{1}{\sqrt{3}}m_{\alpha,\textsf{A}_{2}}+\tfrac{1}{\sqrt{6}}m_{\alpha,\textsf{E}}^{1}+\tfrac{1}{\sqrt{2}}m_{\alpha,\textsf{E}}^{2})],
\end{align}
where on the right side of the equalities the subscript $\alpha=0,1,2,3$ indicates the A-tetrahedron sharing a $0,1,2,3$ site on the B-tetrahedron (Fig.~\ref{fig:convention} in the main text).

When we ignore the heavy modes, in other words, we take $\mathbf{m}_{\text{A},\textsf{T}_{\pm1}}=\mathbf{m}_{\text{A},\textsf{T}_{2}}=0$ on A-tetrahedron, we can represent them in terns of $\mathbb{E}_{\alpha,zz}$ and $\mathbb{E}_{\alpha,zz}^{\pm}$ as follows:
\begin{align}
m_{\text{B},\textsf{A}_{2}}=&\frac{1}{4}\sum_{\alpha=0}^{3}m_{\alpha,\textsf{A}_{2}}=\frac{1}{8\sqrt{3}}\sum_{\alpha=0}^{3}(\mathbb{E}_{\alpha,xx}+\mathbb{E}_{\alpha,yy}+\mathbb{E}_{\alpha,zz})
=\frac{1}{8\sqrt{3}}\sum_{\alpha=0}^{3}(\sqrt{2}(p_{-}\mathbb{E}_{\alpha,zz}^{+}+p_{+}\mathbb{E}_{\alpha,zz}^{-})+\mathbb{E}_{\alpha,zz}),\\
m_{\text{B},\textsf{E}}^{1}=&\frac{1}{4}\sum_{\alpha=0}^{3}m_{\alpha,\textsf{E}}^{1}=\frac{-1}{8\sqrt{6}}\sum_{\alpha=0}^{3}(2\mathbb{E}_{\alpha,xx}-\mathbb{E}_{\alpha,yy}-\mathbb{E}_{\alpha,zz})
=\frac{1}{8\sqrt{6}}\sum_{\alpha=0}^{3}(-\sqrt{5}(p_{\theta}^{+}\mathbb{E}_{\alpha,zz}^{+}+p_{\theta}^{-}\mathbb{E}_{\alpha,zz}^{-})+\mathbb{E}_{\alpha,zz}),\\
m_{\text{B},\textsf{E}}^{2}=&\frac{1}{4}\sum_{\alpha=0}^{3}m_{\alpha,\textsf{E}}^{2}=\frac{-1}{8\sqrt{2}}\sum_{\alpha=0}^{3}(\mathbb{E}_{\alpha,yy}-\mathbb{E}_{\alpha,zz})=\frac{1}{8\sqrt{2}}\sum_{\alpha=0}^{3}(i(\mathbb{E}_{\alpha,zz}^{+}-\mathbb{E}_{\alpha,zz}^{-})+\mathbb{E}_{\alpha,zz}),
\intertext{where $p^{\pm}=e^{\pm\frac{i\pi}{4}}$ and $p_{\theta}^{\pm}=e^{\pm \theta}$ with $\theta=\tan^{-1}(1/2)$, and}
m_{\text{B},\textsf{T}_{1-}}^{x}=&\frac{1}{8}\sum_{\alpha=0}^{3}[2c_{x,\alpha}(\sqrt{\tfrac{2}{3}}m_{\alpha,\textsf{A}_{2}}+\tfrac{1}{\sqrt{3}}m_{\alpha,\textsf{E}}^{1})]
=\frac{1}{8\sqrt{2}}\sum_{\alpha=0}^{3}c_{x,\alpha}[\mathbb{E}_{\alpha,yy}+\mathbb{E}_{\alpha,zz}]\notag\\
=&\frac{1}{8\sqrt{2}}\sum_{\alpha=0}^{3}c_{x,\alpha}[\mathbb{E}_{\alpha,zz}+i(\mathbb{E}_{\alpha,zz}^{-}-\mathbb{E}_{\alpha,zz}^{+})],\\
m_{\text{B},\textsf{T}_{1-}}^{y}=&\frac{1}{8}\sum_{\alpha=0}^{3}[2c_{y,\alpha}(\sqrt{\tfrac{2}{3}}m_{\alpha,\textsf{A}_{2}}-\tfrac{1}{2\sqrt{3}}m_{\alpha,\textsf{E}}^{1}+\tfrac{1}{2}m_{\alpha,\textsf{E}}^{2})]
=\frac{1}{8\sqrt{2}}\sum_{\alpha=0}^{3}c_{x,\alpha}[\mathbb{E}_{\alpha,xx}+\mathbb{E}_{\alpha,zz}]\notag\\
=&\frac{1}{8\sqrt{2}}\sum_{\alpha=0}^{3}c_{y,\alpha}[\mathbb{E}_{\alpha,zz}+(\mathbb{E}_{\alpha,zz}^{-}+\mathbb{E}_{\alpha,zz}^{+})],\\
m_{\text{B},\textsf{T}_{1-}}^{z}=&\frac{1}{8}\sum_{\alpha=0}^{3}[2c_{z,\alpha}(\sqrt{\tfrac{2}{3}}m_{\alpha,\textsf{A}_{2}}-\tfrac{1}{2\sqrt{3}}m_{\alpha,\textsf{E}}^{1}-\tfrac{1}{2}m_{\alpha,\textsf{E}}^{2})]
=\frac{1}{8\sqrt{2}}\sum_{\alpha=0}^{3}c_{x,\alpha}[\mathbb{E}_{\alpha,xx}+\mathbb{E}_{\alpha,yy}]\notag\\
=&\frac{1}{8}\sum_{\alpha=0}^{3}c_{z,\alpha}[p^{+}\mathbb{E}_{\alpha,zz}^{-}+p^{-}\mathbb{E}_{\alpha,zz}^{+}],\\
m_{\text{B},\textsf{T}_{2}}^{x}=&\frac{1}{8}\sum_{\alpha=0}^{3}[-2c_{x,\alpha}m_{\alpha,\textsf{E}}^{2}]
=\frac{1}{8\sqrt{2}}\sum_{\alpha=0}^{3}c_{x,\alpha}[\mathbb{E}_{\alpha,yy}-\mathbb{E}_{\alpha,zz}]
=\frac{1}{8\sqrt{2}}\sum_{\alpha=0}^{3}c_{x,\alpha}[-\mathbb{E}_{\alpha,zz}+i(\mathbb{E}_{\alpha,zz}^{-}-\mathbb{E}_{\alpha,zz}^{+})],\\
m_{\text{B},\textsf{T}_{2}}^{y}=&\frac{1}{8}\sum_{\alpha=0}^{3}[-2c_{y,\alpha}(-\frac{\sqrt{3}}{2}m_{\alpha,\textsf{E}}^{1}-\frac{1}{2}m_{\alpha,\textsf{E}}^{2})]
=\frac{1}{8\sqrt{2}}\sum_{\alpha=0}^{3}c_{y,\alpha}[\mathbb{E}_{\alpha,zz}-\mathbb{E}_{\alpha,xx}]
=\frac{1}{8\sqrt{2}}\sum_{\alpha=0}^{3}c_{y,\alpha}[\mathbb{E}_{\alpha,zz}-(\mathbb{E}_{\alpha,zz}^{-}+\mathbb{E}_{\alpha,zz}^{+})],\\
m_{\text{B},\textsf{T}_{2}}^{z}=&\frac{1}{8}\sum_{\alpha=0}^{3}[-2c_{z,\alpha}(\frac{\sqrt{3}}{2}m_{\alpha,\textsf{E}}^{1}-\frac{1}{2}m_{\alpha,\textsf{E}}^{2})]
=\frac{1}{8\sqrt{2}}\sum_{\alpha=0}^{3}c_{x,\alpha}[\mathbb{E}_{\alpha,xx}-\mathbb{E}_{\alpha,yy}]
=\frac{1}{8}\sum_{\alpha=0}^{3}c_{z,\alpha}[p^{-}\mathbb{E}_{\alpha,zz}^{-}+p^{+}\mathbb{E}_{\alpha,zz}^{+}],\\
m_{\text{B},\textsf{T}_{1+}}^{x}=&-\frac{1}{8}\sum_{\alpha=0}^{3}c_{x,\alpha}[2(\tfrac{1}{\sqrt{3}}m_{\alpha,\textsf{A}_{2}}-\sqrt{\tfrac{2}{3}}m_{\alpha,\textsf{E}}^{1})]
=\frac{-1}{8}\sum_{\alpha=0}^{3}c_{x,\alpha}\mathbb{E}_{\alpha,xx}=\frac{-1}{8}\sum_{\alpha=0}^{3}c_{x,\alpha}(\mathbb{E}_{\alpha,zz}^{-}+\mathbb{E}_{\alpha,zz}^{+}),\\
m_{\text{B},\textsf{T}_{1+}}^{y}=&-\frac{1}{8}\sum_{\alpha=0}^{3}c_{y,\alpha}[
2(\tfrac{1}{\sqrt{3}}m_{\alpha,\textsf{A}_{2}}+\tfrac{1}{\sqrt{6}}m_{\alpha,\textsf{E}}^{1}-\tfrac{1}{\sqrt{2}}m_{\alpha,\textsf{E}}^{2})]
=\frac{-1}{8}\sum_{\alpha=0}^{3}c_{y,\alpha}\mathbb{E}_{\alpha,yy}=\frac{-1}{8}\sum_{\alpha=0}^{3}ic_{y,\alpha}(\mathbb{E}_{\alpha,zz}^{-}-\mathbb{E}_{\alpha,zz}^{+}),\\
m_{\text{B},\textsf{T}_{1+}}^{z}=&-\frac{1}{8}\sum_{\alpha=0}^{3}c_{z,\alpha}[
2(\tfrac{1}{\sqrt{3}}m_{\alpha,\textsf{A}_{2}}+\tfrac{1}{\sqrt{6}}m_{\alpha,\textsf{E}}^{1}+\tfrac{1}{\sqrt{2}}m_{\alpha,\textsf{E}}^{2})]=\frac{-1}{8}\sum_{\alpha=0}^{3}c_{z,\alpha}\mathbb{E}_{\alpha,zz}.
\end{align}

\end{widetext}

\begin{figure}
\centering
\includegraphics{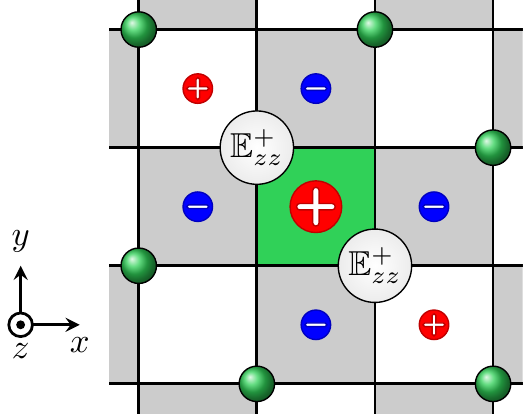}
\caption{The top-down view of charge configuration for $\mathbb{E}_{1,zz}^{+}\mathbb{E}_{2,zz}^{+}$. Here, the gray and white squares represent the positions of the charges located on $z=1/4$ and $z=-1/4$, respectively. The green circles indicate the location of the A-tetrahedra on the same $xy$-plane, $z=0$. In contrast with $\mathbb{E}_{1,zz}^{-}\mathbb{E}_{2,zz}^{+}$, the charge on the overlapped region (green square) between the charge configurations created by $\mathbb{E}_{1,zz}^{+}$ and $\mathbb{E}_{2,zz}^{+}$ is not cancelled out, but piled up, so there is $+2$ $z$-charge represented by a big red circle with $+$ sign.}\label{fig:charge_pp}
\end{figure}

\begin{figure*}
\includegraphics{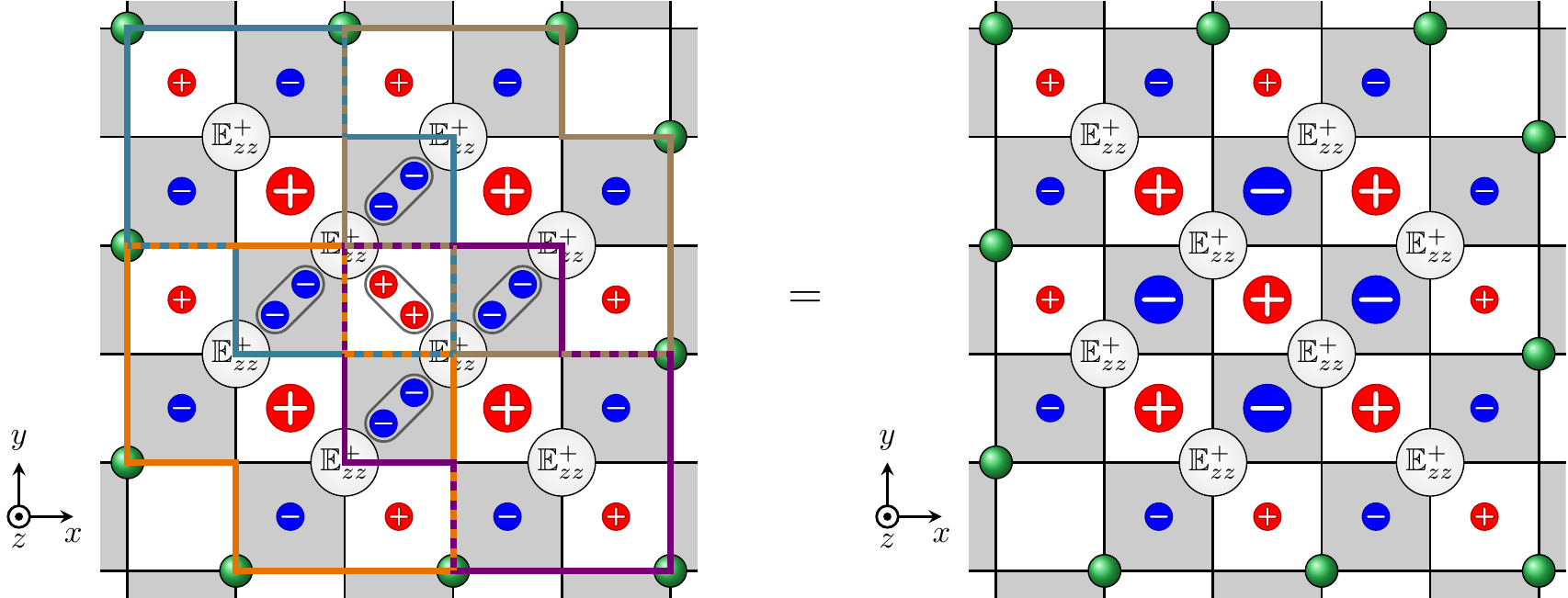}
\caption{Depiction of the fourth-order perturbation by the membrane operators consisting of $\mathbb{E}_{1,zz}^{+}\mathbb{E}_{2,zz}^{+}$. The red circle with $+$ sign and blue circle with $-$ sign stand for the $\pm1$ $z$-charges, and the big red circle with $+$ sign stands for $+2$ $z$-charge, respectively. The charges in the overlapped region are piled up, so $\pm2$ $z$-charges are inside of it and $\pm1$ $z$-charges on the edge.}\label{fig:bulk_perturb_4}
\end{figure*}

\begin{figure*}
\includegraphics{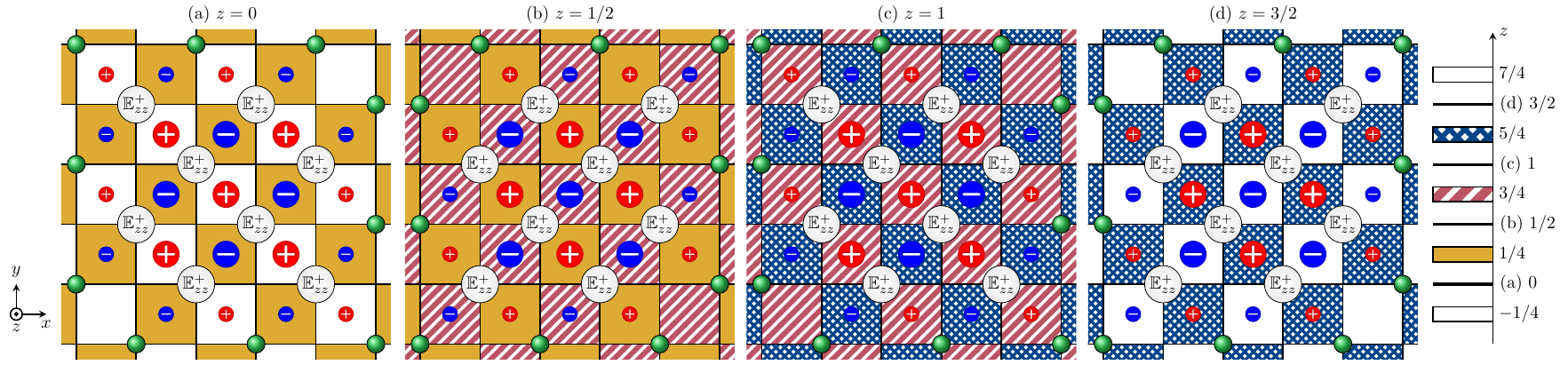}
\caption{\color{black}
Depiction of stacking of the perturbations for different sets of operators for $L_{x}=L_{y}=L_{z}=2$. 
(a-d) The charge configurations when we apply the membrane operators consisting of $\mathbb{E}_{1,zz}^{+}\mathbb{E}_{2,zz}^{+}$ and $\mathbb{E}_{0,zz}^{+}\mathbb{E}_{3,zz}^{+}$ on $xy$ planes at $z=0$, $1/2$, $1$, and $3/2$, respectively.
In other words, we apply the raising operators $\mathbb{E}_{zz}^{+}$ on the A-tetrahedron sites on the $xy$ planes at $z=0$, $1/2$, $1$, and $3/2$, respectively. 
At this time, $z$-charges are created above and below the plane.
The white, yellow, red-diagonal, and blue-diagonal-grid squares stand for the positions at which the $z$-charges are created when we apply the membrane operators, and they are located on $z=-1/4$, $1/4$, $3/4$, and $5/4$, respectively. 
The red circle with $+$ sign and blue circle with $-$ sign stand for the $\pm1$ $z$-charge, and
the big red circle with $+$ sign and big blue circle with $-$ sign stand for the $\pm2$ $z$-charge, respectively. 
For example, if we apply the membrane operator on $xy$ plane at $z=0$ (charge configuration in (a)), $-2$ $z$-charges are created at $z=1/4$ (yellow squares in (a)) and $+2$ $z$-charges are created at $z=-1/4$ (white squares in (a)).
The charges on overlapped regions between adjacent $xy$ planes are cancelled out. For example, in (a) and (b), the yellow squares are overlapped because they are located at same $z=1/4$ position, and the charges on the overlapped region are cancelled out because the charges from each configuration have opposite sign. Still remaining are the charges on the corners.
As a result, by stacking the charge configurations by the membrane operators consisting of  $\mathbb{E}_{1,zz}^{+}\mathbb{E}_{2,zz}^{+}$ and $\mathbb{E}_{0,zz}^{+}\mathbb{E}_{3,zz}^{+}$ on $xy$ planes, we can get the boundary charges on the bottom ($z=-1/4$) and top ($z=7/4$) surfaces, and the hinges (corners).
The right bar legend stand for the locations at which the membrane operators ($\mathbb{E}_{zz}^{+}$) are applied and $\pm2$ charges are created in terms of $z$ positions.
}\label{fig:bulk_perturb_bulk}
\end{figure*}

\section{Boundary charges from other perturbative terms}\label{app:bulk_op}

\begin{table*}
\begin{tabular}{|>{$}c<{$}|>{$}c<{$}|>{$}c<{$}|}
\hline
\text{planes}&\text{stacked operators}&\text{membrane replacements}\\
\hline\hline
xy&\mathbb{E}_{1,zz}^{\pm}\mathbb{E}_{2,zz}^{\pm}\text{ and }\mathbb{E}_{0,zz}^{\pm}\mathbb{E}_{3,zz}^{\pm}&\mathbb{E}_{0,zz}^{\pm}\mathbb{E}_{2,zz}^{\pm}\text{ and }\mathbb{E}_{1,zz}^{\pm}\mathbb{E}_{3,zz}^{\pm}\text{ on $xz$ planes}/\mathbb{E}_{0,zz}^{\pm}\mathbb{E}_{1,zz}^{\pm}\text{ and }\mathbb{E}_{2,zz}^{\pm}\mathbb{E}_{3,zz}^{\pm}\text{ on $yz$ planes}\\
\hline
xz&\mathbb{E}_{0,zz}^{\pm}\mathbb{E}_{2,zz}^{\mp}\text{ and }\mathbb{E}_{1,zz}^{\pm}\mathbb{E}_{3,zz}^{\mp}&\mathbb{E}_{1,zz}^{\pm}\mathbb{E}_{2,zz}^{\mp}\text{ and }\mathbb{E}_{0,zz}^{\pm}\mathbb{E}_{3,zz}^{\mp}\text{ on $xy$ planes}/\mathbb{E}_{0,zz}^{\pm}\mathbb{E}_{1,zz}^{\pm}\text{ and }\mathbb{E}_{2,zz}^{\mp}\mathbb{E}_{3,zz}^{\mp}\text{ on $yz$ planes}\\
\hline
yz&\mathbb{E}_{0,zz}^{\pm}\mathbb{E}_{1,zz}^{\mp}\text{ and }\mathbb{E}_{2,zz}^{\pm}\mathbb{E}_{3,zz}^{\mp}&\mathbb{E}_{1,zz}^{\mp}\mathbb{E}_{2,zz}^{\pm}\text{ and }\mathbb{E}_{0,zz}^{\pm}\mathbb{E}_{3,zz}^{\mp}\text{ on $xy$ planes}/\mathbb{E}_{0,zz}^{\pm}\mathbb{E}_{2,zz}^{\pm}\text{ and }\mathbb{E}_{1,zz}^{\mp}\mathbb{E}_{3,zz}^{\mp}\text{ on $xz$ planes}\\
\hline
\end{tabular}
\caption{The stacked operators and their replacements by the combination of the membrane operators.
For example, if we apply the membrane operators consisting of $\mathbb{E}_{1,zz}^{\pm}\mathbb{E}_{2,zz}^{\pm}\text{ and }\mathbb{E}_{0,zz}^{\pm}\mathbb{E}_{3,zz}^{\pm}$ on all $xy$ planes, we can get the stacked operators. And the stacked operators can be replaced by the combinations of the membrane operators consisting of $\mathbb{E}_{0,zz}^{\pm}\mathbb{E}_{2,zz}^{\pm}\text{ and }\mathbb{E}_{1,zz}^{\pm}\mathbb{E}_{3,zz}^{\pm}$ on $xz$ planes, or $\mathbb{E}_{0,zz}^{\pm}\mathbb{E}_{1,zz}^{\pm}\text{ and }\mathbb{E}_{2,zz}^{\pm}\mathbb{E}_{3,zz}^{\pm}$ on $yz$ planes.
}\label{tab:bulk}
\end{table*}

\begin{figure*}
\subfigure[]{
\includegraphics[height=19em]{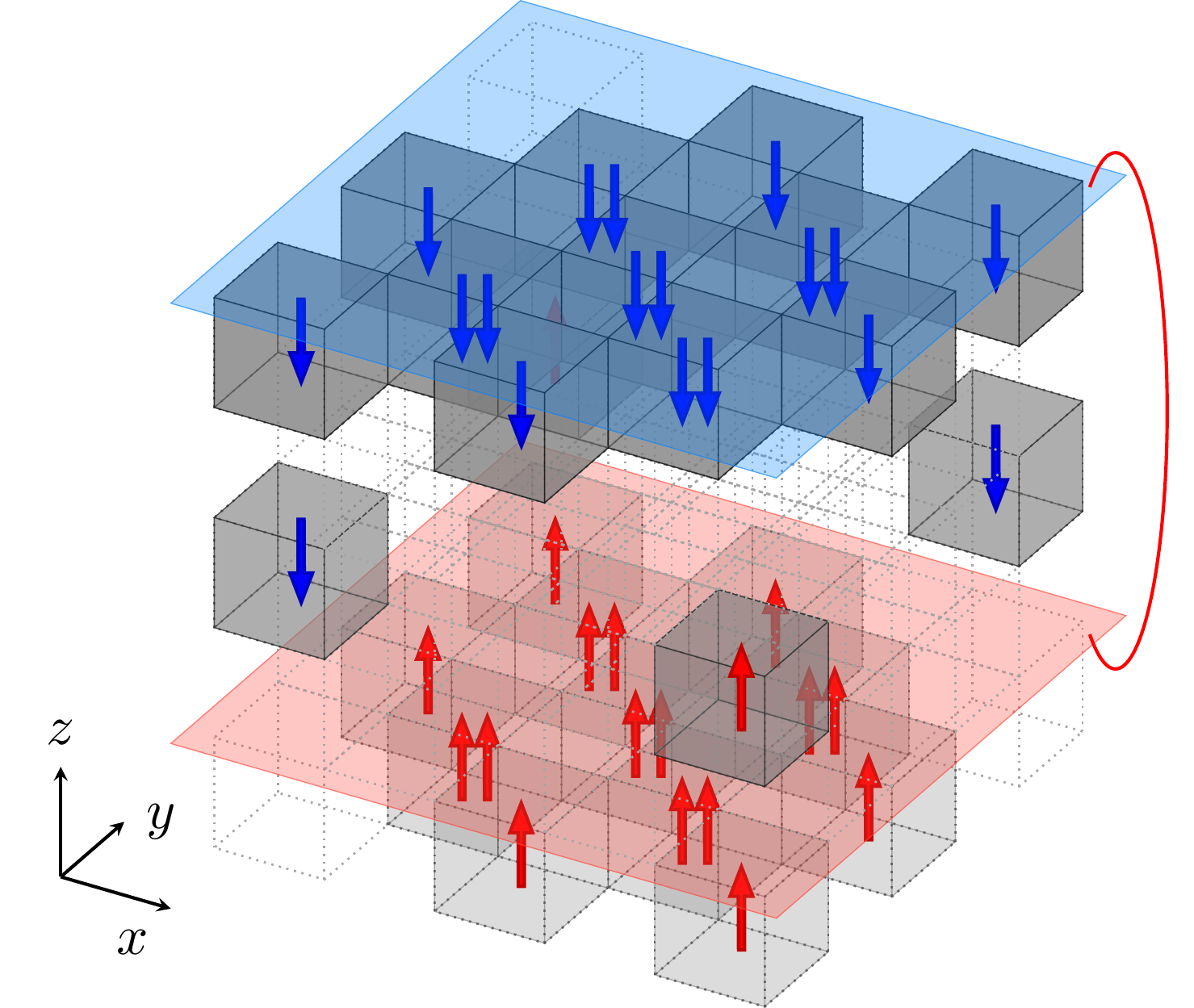}
\label{fig:bulk_stack}}
\subfigure[]{
\includegraphics[height=19em]{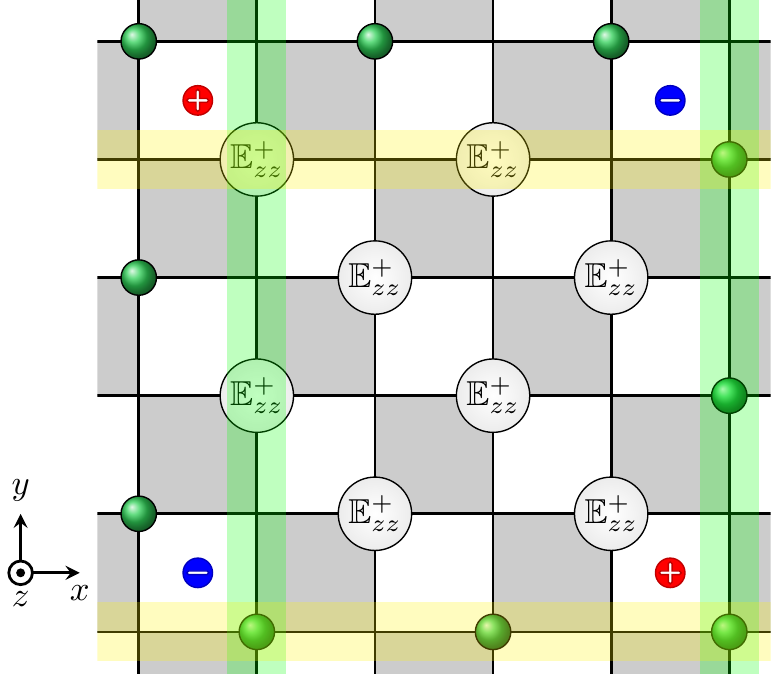}
\label{fig:bulk_pbc}}
\caption{The stacking and periodic boundary conditions of the resulting charge configuration by the membrane operators consisting of $\mathbb{E}_{1,zz}^{+}\mathbb{E}_{2,zz}^{+}$ when $L_{x}=L_{y}=2$. 
The light gray, gray, and dark gray cubics represent the positions of the charges located at $z=-1/4$, $3/4$, and $7/4$, respectively.
(a) The stacking of the charge configuration by the membrane operators consisting of $\mathbb{E}_{1,zz}^{+}\mathbb{E}_{2,zz}^{+}$ when $L_{x}=L_{y}=2$. The red upward and blue downward arrows mean $\pm1$ $z$-charges, respectively.
The bundles of two red or two blue arrows stands for $\pm2$ $z$-charges, respectively
By stacking the charge configurations in Fig.~\ref{fig:bulk_perturb_bulk}, we get the boundary charges on bottom ($z=-1/4$) and top ($z=7/4$) surfaces, and hinges ($z=3/4$), respectively. The red and blue planes are located at $z=0$ and $z=2$, respectively.
By identifying the red and blue planes, the charges on the overlapped region between bottom and top surfaces are cancelled out, and the hinge charges on corners at $z=3/4$ and $z=7/4$ remain.
(b) The top-down view of the hinge (corner) charge configuration after taking the periodic boundary condition on $z$ direction.
By identifying the green and yellow lines, the hinge (corner) charges are cancelled out, we can get the charge-neutral vacuum.
}\label{fig:bulk_stack_pbc}
\end{figure*}

Here, we will discuss the possible perturbations from the other perturbation terms,
described in the main text.
For illustration, let us consider the same example with the main text, which is $\mathbb{E}_{1,zz}^{+}\mathbb{E}_{2,zz}^{+}$ on the $xy$ plane. 
At the first-order perturbation, it results in a state that has the charge configuration presented in Fig.~\ref{fig:charge_pp}. 
In contrast with the case of $\mathbb{E}_{1,zz}^{+}\mathbb{E}_{2,zz}^{-}$, the charge on the overlapped region (represented by green square in Fig.~\ref{fig:charge_pp}) between the charge configurations created by $\mathbb{E}_{1,zz}^{+}$ and $\mathbb{E}_{2,zz}^{+}$ is not cancelled out, but piled up, so there is $+2$ $z$-charge represented by a big red circle with $+$ sign.
If we try to perform the higher-order perturbation by using $\mathbb{E}_{1,zz}^{+}\mathbb{E}_{2,zz}^{+}$ on $xy$ plane in the similar way as the membrane operator composed of $\mathbb{E}_{1,zz}^{+}\mathbb{E}_{2,zz}^{-}$ in the main text, 
the charges on the overlapped region are not cancelled, but they are accumulated. As a result, the resulting state has $\pm2$ $z$-charges inside of it (Fig.~\ref{fig:bulk_perturb_4}).
Even if we take the periodic boundary condition on $x$ and $y$ directions, the plane is covered by $\pm2$ $z$-charges. 
As such, we cannot return back to the charge-neutral ground state by using the membrane operator of $\mathbb{E}_{1,zz}^{+}\mathbb{E}_{2,zz}^{+}$, unlike the membrane operator of $\mathbb{E}_{1,zz}^{+}\mathbb{E}_{2,zz}^{-}$. 
However, there is a way to return back to the charge-neutral vacuum as follows. We need to apply the membrane operator consisting of $\mathbb{E}_{1,zz}^{+}\mathbb{E}_{2,zz}^{+}$ and $\mathbb{E}_{0,zz}^{+}\mathbb{E}_{3,zz}^{+}$ on all the other  (stacked) planes (Fig.~\ref{fig:bulk_perturb_bulk}). 
For example, let us consider the charge configurations (a) and (b) in Fig.~\ref{fig:bulk_perturb_bulk} created by applying the membrane operators consisting of $\mathbb{E}_{1,zz}^{+}\mathbb{E}_{2,zz}^{+}$ and $\mathbb{E}_{0,zz}^{+}\mathbb{E}_{3,zz}^{+}$ on $xy$ planes located $z=0$ and $z=1/2$, respectively.
Then, the charges located at the overlapped regions (yellow squares located at $z=1/4$) between two charge configurations are cancelled out, and there remain alternative $\pm1$ $z$-charges at the corners.
If we keep repeating this on the other planes in $z$ direction, we can cancel out the bulk charges inside of it (Fig.~\ref{fig:bulk_stack}).
As a result, we will get the remaining boundary charges on the bottom and top surfaces, and hinges (corners). If we take the periodic boundary condition on $z$ direction, we can cancel out the boundary charges on the bottom and top surfaces, and we have the remaining charges at the corner (hinges). These charges on the hinges can be cancelled out by taking the periodic boundary conditions on $x$ or $y$ directions.
For example, in Fig.~\ref{fig:bulk_stack}, we stack the charge configurations in Fig.~\ref{fig:bulk_perturb_bulk} generated by the membrane operators consisting of $\mathbb{E}_{1,zz}^{+}\mathbb{E}_{2,zz}^{+}$ and $\mathbb{E}_{0,zz}^{+}\mathbb{E}_{3,zz}^{+}$, so we have the boundary charges on the bottom ($z=-1/4$) and top ($z=7/4$) surfaces, and hinges.
If we identify the blue ($z=0$) and red ($z=2$) planes, the charges below the blue and the red planes are cancelled out, and there remain the charges at the corners (Fig.~\ref{fig:bulk_pbc}). So we have the charges at the corners (hinges) located at $z=3/4$ and $z=7/4$ ($z=7/4$ is the same with $z=-1/4$ because we now identify $z=0$ and $z=2$). 
As a result, $\mathbb{E}_{1,zz}^{+}\mathbb{E}_{2,zz}^{+}$ and $\mathbb{E}_{0,zz}^{+}\mathbb{E}_{3,zz}^{+}$ generate the aforementioned perturbations to return back to the charge-neutral vacuum.

\begin{figure*}
\includegraphics[width=\textwidth]{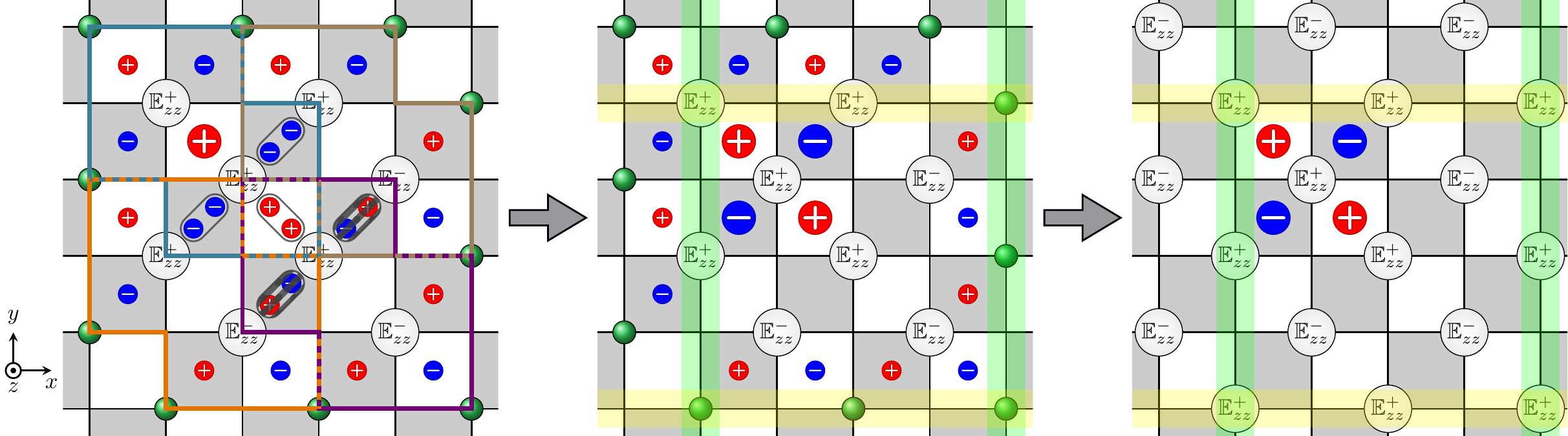}
\caption{
The depiction of the mixture of the operators. We can get the figure on the left panel if we replace one $\mathbb{E}_{1,zz}^{+}\mathbb{E}_{2,zz}^{-}$ by $\mathbb{E}_{1,zz}^{+}\mathbb{E}_{2,zz}^{-}$ in Fig.~\ref{fig:perturb_4}. If we take the perturbation on the left panel, we can get the charge configuration on the middle panel. In the overlapped regions, the charges have the opposite sign, then they are cancel out. However, if they has the same sign, they are piled up.
The red circle with $+$ sign and blue circle with $-$ sign stand for the $\pm1$ $z$-charge, and
the big red circle with $+$ sign and big blue circle with $-$ sign stand for the $\pm2$ $z$-charge, respectively. 
On the middle panel, by taking the periodic boundary condition on the $x$ and $y$ directions, we get a cluster of the charges on the top left-hand corner (right panel). The green and yellow lines are identified, respectively.
}\label{fig:mix_op}
\end{figure*}

Note that these stacked operators can be replaced by the other membrane operators introduced in the main text. For the above example, the stacked operators increase $\mathbb{E}_{zz}$ quantum numbers by 1 at all the A-tetrahedra. It means that the stacked operator consisting of $\mathbb{E}_{1,zz}^{+}\mathbb{E}_{2,zz}^{+}$ and $\mathbb{E}_{0,zz}^{+}\mathbb{E}_{3,zz}^{+}$ can be replaced by the combinations of the membrane operators consisting of $\mathbb{E}_{0,zz}^{+}\mathbb{E}_{2,zz}^{+}$ and $\mathbb{E}_{1,zz}^{+}\mathbb{E}_{3,zz}^{+}$ on $xz$ planes, or $\mathbb{E}_{0,zz}^{+}\mathbb{E}_{1,zz}^{+}$ and $\mathbb{E}_{2,zz}^{+}\mathbb{E}_{3,zz}^{+}$ on $yz$ planes, which are introduced in the main text, respectively. 
It is the same for all other combinations of the operators that do not make use of the membrane operators in the main text.
Thus, all the perturbations that tunnel between the ground states can be expressed in terms of the membrane operators which are introduced in the main text, and as such the membrane operators can be regarded as the fundamental operators on the breathing pyrochlore lattice.
We present the summary for the stacked operators and their membrane replacements in Table~\ref{tab:bulk}.

\section{Mixture of perturbative terms}\label{app:mix_op}
In the main text, we briefly discussed that a mixture of the perturbative operators prohibits a tunnelling process between the ground states. Here, we will present an illustrative example about it.
Consider Fig.~\ref{fig:mix_op}, where we replace one $\mathbb{E}_{1,zz}^{+}\mathbb{E}_{2,zz}^{-}$ by $\mathbb{E}_{1,zz}^{+}\mathbb{E}_{2,zz}^{+}$ in Fig.~\ref{fig:perturb_4}.
After taking periodic boundary conditions, we arrive at a cluster of charges on the top left-hand corner. Since this charge configuration is centered at the A-tetrahedron site, we cannot easily cancel it out by creating charge configurations at the adjacent $xy$ planes because the A-tetrahedron sites on the adjacent planes are not located directly above in the face-centered cubic lattice. Due to this geometric mismatch, charges on adjacent planes are unable to be cancelled out in the manner of Appendix~\ref{app:bulk_op}.
As we discussed in Sec.~\ref{sec:absence_mag} of the main text, in our lattice geometry, since the charges are created tetragonally in three-dimension, the best ways to cancel out the charges are the membrane operators or stacked operators of Appendix~\ref{app:bulk_op}. 
For that reason, the mixture of the operators cannot tunnel between the ground states, but can make charge-ful excited states.

\section{Role of diagonal perturbative terms}
\label{app_diagonal_terms_perturb}
In the main text, we discussed in detail the role of possible perturbative terms that allow the system to tunnel between the various ground state manifolds. 
This was motivated by earlier works in quantum spin ice, where transverse coupling terms allowed the tunnelling between the largely degenerate two-in, two-out ice states, which manifested a ring-exchange magnetic vector potential.
Though understanding the delicate stability of the proposed quantum fractonic ground state (in particular, whether it may survive when considered along with the charge-creating perturbative terms) requires a more elaborate numerical study, we can nonetheless glean the effects of such a term by simply considering the possible energy shift it may have on our ground state manifold within first-order perturbation theory.
 
Let us consider the expectation value for $\mathbb{E}_{A,zz}\mathbb{E}_{A',zz}$ in terms of the ground states. The expectation value is given by
\begin{align}
\notag\langle\sum_{A,A'}a_{AA'}&\hat{\mathbb{E}}_{A,zz}\hat{\mathbb{E}}_{A',zz}\rangle{}_{\psi}\\
={}&\braket{\psi|\sum_{A,A'}a_{AA'}\hat{\mathbb{E}}_{A,zz}\hat{\mathbb{E}}_{A',zz}|\psi}\notag\\
={}&-\frac{(J_B-K_B)}{64}\sum_{\text{B}} \Big(\sum_{\text{B}_{\alpha}}\mathbb{E}_{\text{B}_{\alpha},zz}\Big)^{2}\nonumber \\
 &-\frac{(J_B+K_B)}{64}\sum_{\text{B}} \Big(\sum_{\text{B}_{\alpha}}c_{x,\text{B}_{\alpha}}\mathbb{E}_{\text{B}_{\alpha},zz}\Big)^{2} \nonumber \\
 &-\frac{(J_B+K_B)}{64}\sum_{\text{B}} \Big(\sum_{\text{B}_{\alpha}}c_{y,\text{B}_{\alpha}}\mathbb{E}_{\text{B}_{\alpha},zz}\Big)^{2}\label{eq:EzzEzz}
\end{align}
where $\psi$ stands for a given ground state, $\hat{\mathbb{E}}_{\alpha,zz}$ and $\mathbb{E}_{\alpha,zz}$ are the electric field quantum number operator and eigenvalue at $\alpha$, respectively. $B_{\alpha}$ stands for the $\alpha$-th $A$-tetrahedron surrounding $B$-tetrahedron. The coefficients $a_{AA'}$ are determined by Eq.~(D13-D24) in Appendix D. As is evident, the energy shift due to this term depends on the Heisenberg $J_B$ and $K_B$ from sublattice B. 

\begin{figure*}[t]
\includegraphics[width=0.9\textwidth]{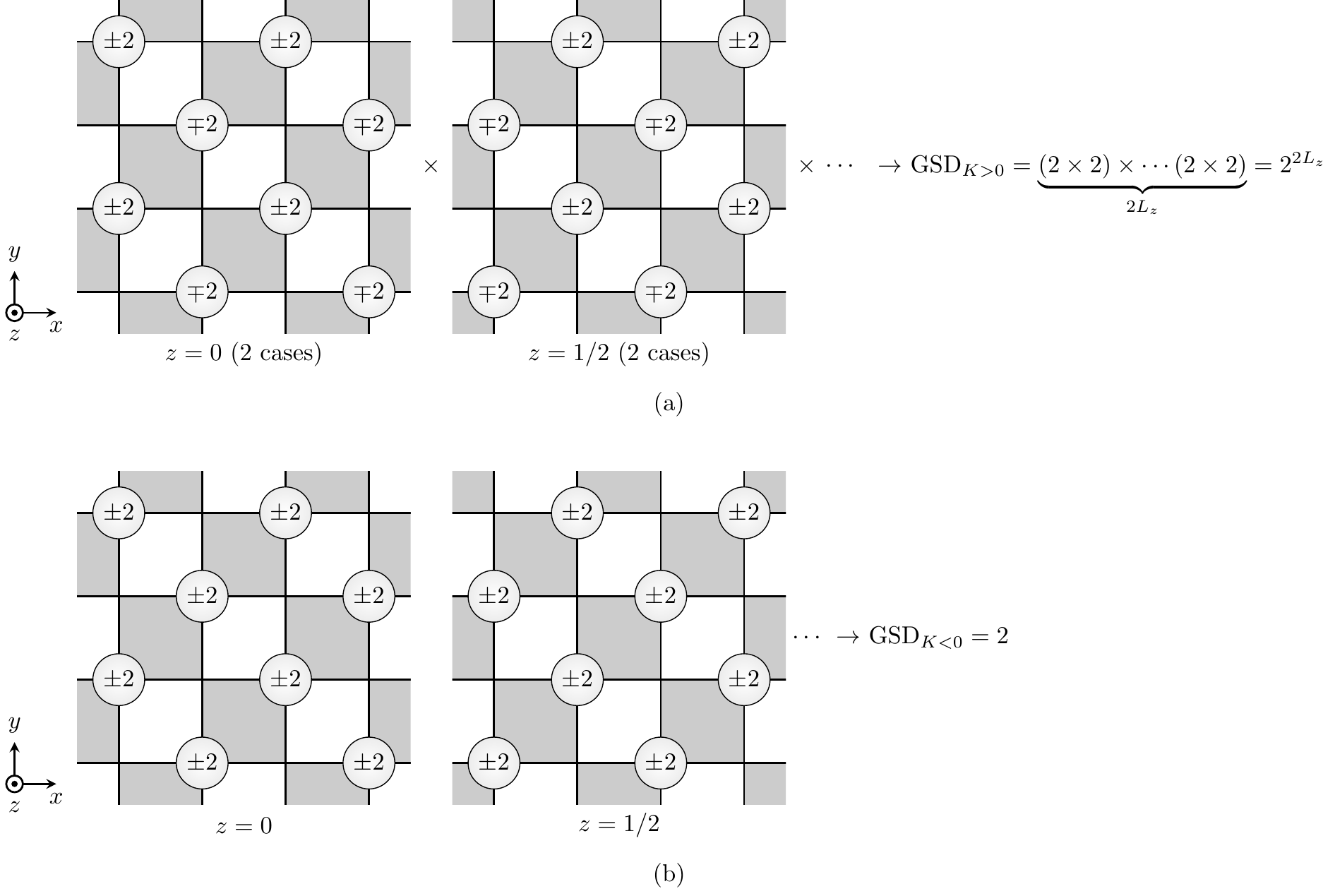}
\caption{The electric field quantum number configurations of the remnant ground states for (a) $K>0$ and (b) $K<0$.}
\label{fig:EzzEzz}
\end{figure*}
\begin{table}[h!]
\begin{tabular}{>{$}c<{$}|>{$}c<{$}|>{$}c<{$}|>{$}r<{$}|>{$}c<{$}|>{$}c<{$}|>{$}c<{$}}\hline
L_{x}&L_{y}&L_{z}&\text{GSD}&\text{GSD}_{K=0}&\text{GSD}_{K>0}&\text{GSD}_{K<0}\\
\hline\hline
1&1&1&85&6&4&2\\
\hline
2&1&1&1,333&18&4&2\\
1&1&2&1,333&18&16&2\\
\hline
3&1&1&25,405&66&4&2\\
1&1&3&25,405&66&64&2\\
\hline
4&1&1&535,333&258&4&2\\
1&1&4&535,333&258&256&2\\
\hline
2&2&1&10,213&30&4&2\\
1&2&2&10,213&30&16&2\\
\hline
2&2&2&49,541&42&16&2\\
\hline
3&2&1&116,653&78&4&2\\
1&3&2&116,653&78&16&2\\
1&2&3&116,653&78&64&2\\
\hline
\end{tabular}
\caption{Table for the ground state degeneracy. GSD and GSD$_{K\gtreqqless0}$ stand for the ground state degeneracies in the absence and presence of $\mathbb{E}_{A,zz}\mathbb{E}_{A',zz}$, respectively.}\label{tab:remnant}
\end{table}

In our work, motivated by concrete material examples such as Ba$_3$Yb$_2$Zn$_5$O$_{11}$ \cite{PhysRevB.93.220407,PhysRevLett.116.257204} we take $J$ is positive on both of A- and B-sublattices. 
As such, the fate of the ground state degeneracy due to the above perturbative term rest on $K_B$. 
When $K_B=0$, the electric field quantum number configurations of the remnant ground states consist of $|\mathbb{E}_{A,zz}|=2$ at each A-tetrahedron.
When $K_B>0$, the second and third terms in Eq.~\ref{eq:EzzEzz} will play a dominant role. 
In this case, on each $xy$ plane, the electric field quantum numbers on the line along a $y$-direction are the same, while the adjacent lines have opposite signs (see Fig.~\ref{fig:EzzEzz}(a)).
So, for each $xy$ plane, we have two possibilities and since we have $2L_{z}$ number of $xy$ planes, the resulting remnant ground state degeneracy is $\text{GSD}_{K_B>0}=(2\times2)\times\cdots\times(2\times2)=2^{2L_{z}}$. 
On the other hand, when $K_B<0$, the first term in Eq.~\ref{eq:EzzEzz} will play a dominant role. 
Indeed, it is accomplished by configurations where all the electric field quantum numbers are equal to $+2$ or $-2$ at all of A-tetrahedra (see Fig.~\ref{fig:EzzEzz}). 
For example, in the $(1,1,1)$ geometry, these states correspond to the $(\mathbb{E}_{0,zz},\mathbb{E}_{1,zz},\mathbb{E}_{2,zz},\mathbb{E}_{3,zz})=(2,2,2,2)$ and $(-2,-2,-2,-2)$ configurations.
We list the remnant ground state degeneracies for a number of geometries Table~\ref{tab:remnant}.

Generalizing the results from Table~\ref{tab:remnant}, we find that the remnant GSD with $K_B=0$ for given lattice geometry is given by the simple analytic form $\text{GSD}_{K_B=0}=2^{2L_{1}}+2+12(L_{2}-1)+12(L_{3}-1)=2^{2L_{1}}+12(L_{2}+L_{3})+22$ where $L_{1}$ is a maximum value among $L_{x}$, $L_{y}$, and $L_{z}$, and $L_{2}$ and $L_{3}$ are second and third highest values, respectively. 
The remnant GSD with $K_B>0$ is $2^{2L_{z}}$, while the remnant GSD $K_B<0$ is $2$, as mentioned before. 
This implies that in the thermodynamic limit, $L_{x,y,z}\rightarrow\infty$, the ground state degeneracy 
for $K_B \geq 0$ is dependent on lattice geometry, exponentially grows with system size, and is sub-extensive in system volume.
As such, even with the inclusion of such a term, we are led to find that the phase of matter still possesses characteristics of quantum fractonic ground states.

\section{Geometrical restrictions prohibiting finite-order perturbative processes}
\label{app_mag_field_none}

\begin{figure}
\subfigure[]{
\includegraphics[width=0.3\linewidth]{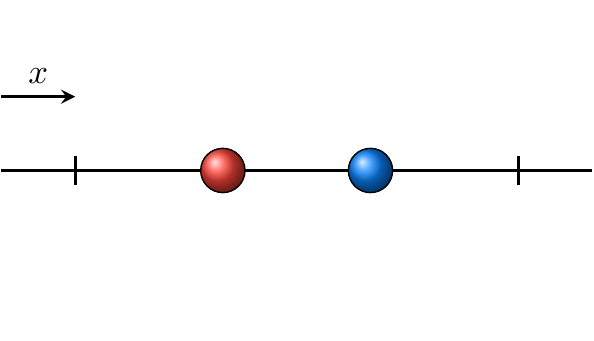}\label{fig:one_dim_a}}
\subfigure[]{
\includegraphics[width=0.3\linewidth]{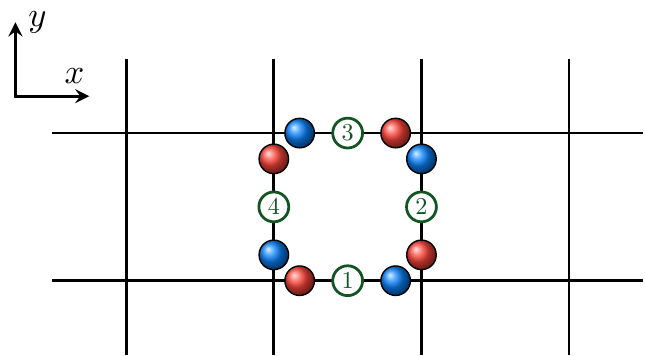}\label{fig:one_dim_b}}\\
\subfigure[]{
\includegraphics[width=0.3\linewidth]{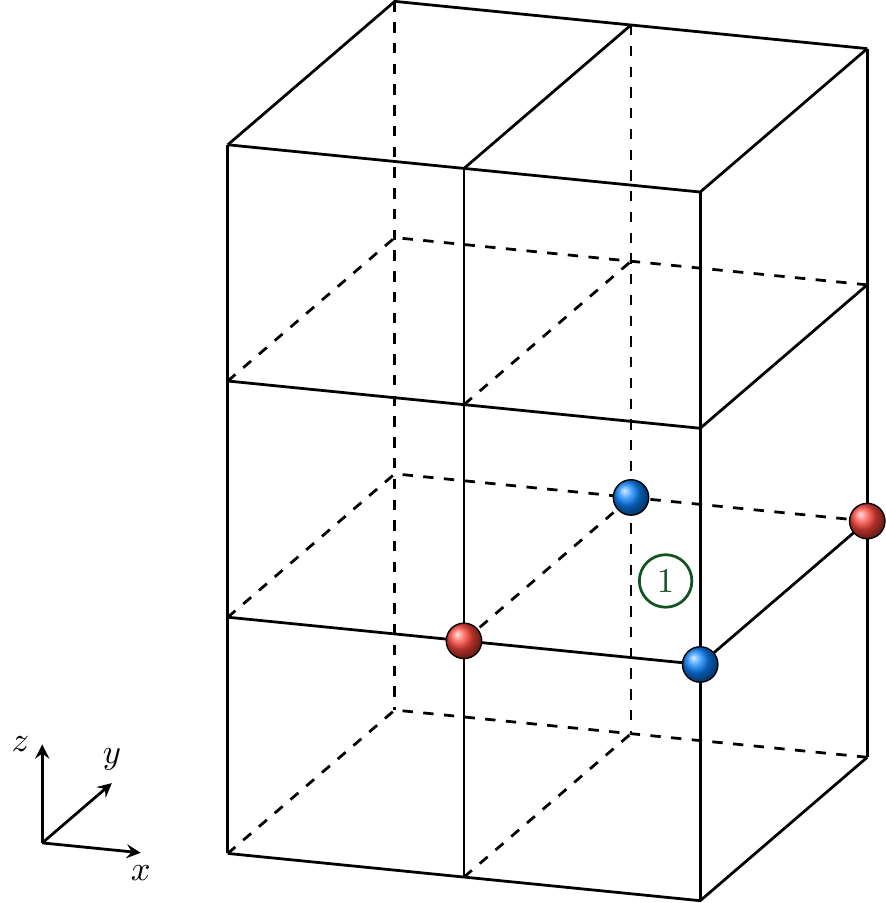}\label{fig:two_dim_a}}
\subfigure[]{
\includegraphics[width=0.3\linewidth]{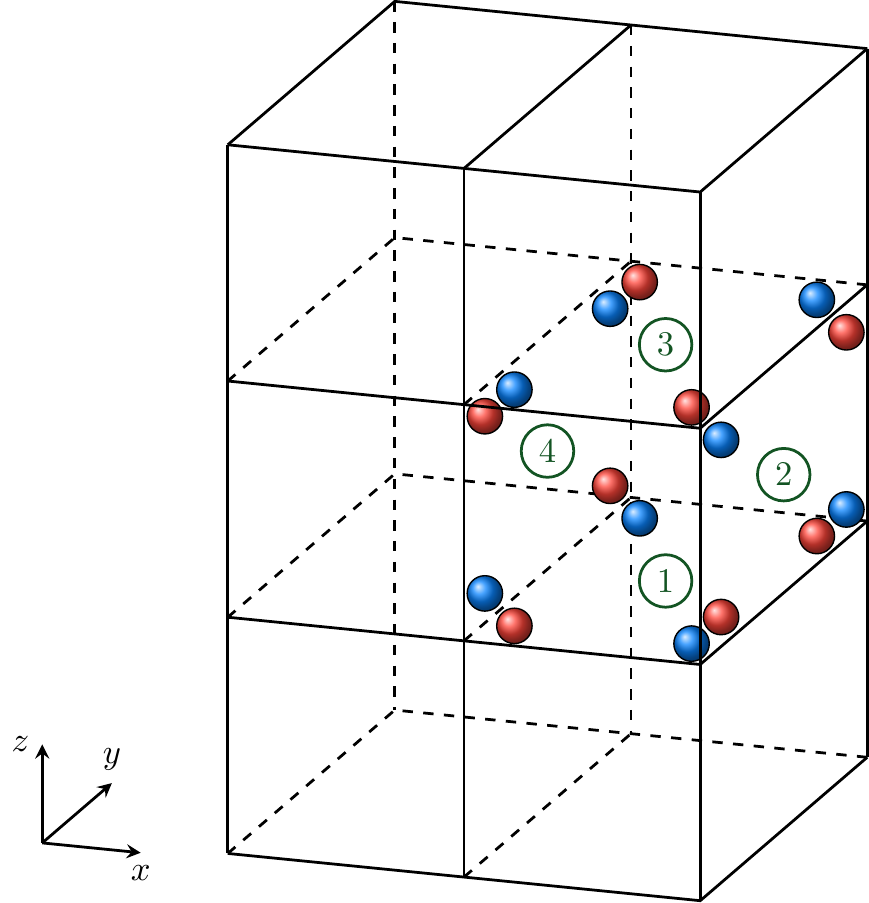}\label{fig:two_dim_b}}\\
\subfigure[]{
\includegraphics[width=0.3\linewidth]{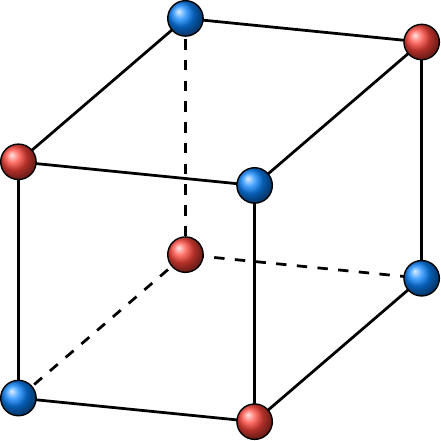}\label{fig:three_dim_a}}
\subfigure[]{
\includegraphics[width=0.3\linewidth]{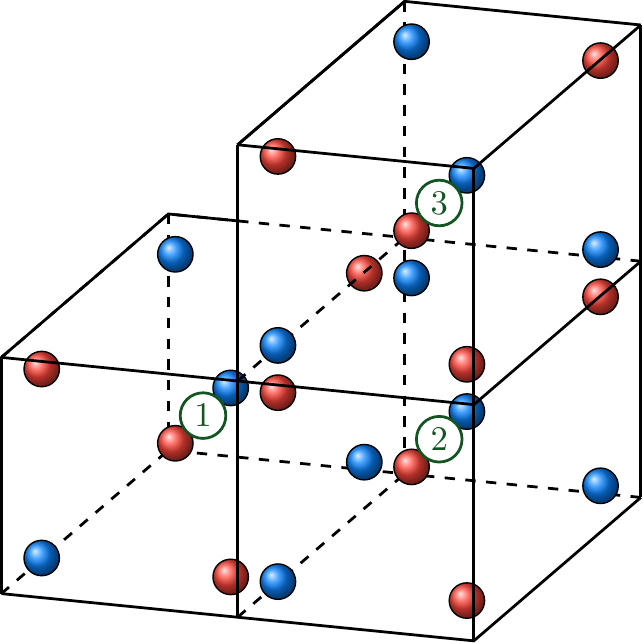}\label{fig:three_dim_b}}
\caption{Degree of complexity of eliminating corner edge charges in different dimensions on a simple cubic lattice. (a)-(b) One-dimensional corner charge elimination by utilizing two-dimensional pathway. (c)-(d) Two-dimensional corner charge elimination by utilizing three-dimensional pathway. (e)-(f) Inability to eliminate three-dimensional corner charges using any three-dimensional pathways. }
\label{fig:charge_lower_dim}
\end{figure}

The primary source of difficulty in the generation of a finite-order magnetic field is the complicated three-dimensional geometry of the breathing pyrochlore lattice.
The application of a raising/lowering operator leads to charges being created in a three-dimensional volume as seen in Fig.~\ref{fig:charge_config}.
This is unlike the case of creating charges in a (lower-dimensional) line or plane where a finite-order perturbation can be more easily realized.
Consider the scenario where charges are created along a one-dimensional line, as seen in Fig.~\ref{fig:one_dim_a} by acting a raising/lowering operator of a link connecting two sites.
Applying raising/lowering operators parallel to the line results in edges being created at the ends of the line.
To circumvent this `corner edge' problem, one can consider applying the operator along a perpendicular direction (to the line of charge) that allows the charges to ``wrapped around'' and eventually cancel each other (as shown in Fig.~\ref{fig:one_dim_b}); in this example, each site has an equal number of positive (red) and negative (blue) charges.
As such, by employing a higher dimension, one can imagine the generation of a magnetic field at finite order in perturbation.
This one-dimensional ``line of charge'' construction is a simple way to understand why a finite-order perturbation was permitted in the quantum spin ice setting (in that case the charges are wrapped around a hexagon \cite{PhysRevB.69.064404}).
Next consider charges created in a two-dimensional plane of a cubic lattice in Fig.~\ref{fig:two_dim_a} by acting a raising/lowering operator at the centre of a square face.
Once again, corner charges on a plane are created if the raising/lowering operators are continuously applied parallel to the plane of charges.
The charges can be eliminated by once again accessing a higher dimension: in this example, perpendicular to the plane of charges as seen in Fig.~\ref{fig:two_dim_b} such that each site has an equal number of positive (red) and negative (blue) charges.
Through these lower-dimensional examples, one can appreciate the difficulty in eliminating the charges at finite order in perturbation as seen in a simple three-dimensional example of Fig.~\ref{fig:three_dim_a} and \ref{fig:three_dim_b}.
Even in this depicted ``simpler'' geometry of a simple cubic lattice the corner charges cannot be easily eliminated by raising/lowering operators acting at the centre of the cubes; as seen, each site has an inequal number of positive (red) and negative (blue) charges.
Indeed, by extending the ideas of eliminating lower-dimensional charge configurations, it suggests that an additional (and not achievable in this setting) fourth dimension may be required to eliminate the corner charge on the breathing pyrochlore lattice.

%

\end{document}